\documentclass[acmtog]{acmart} 
\acmSubmissionID{414}

\usepackage{booktabs} 
\usepackage{enumitem}
\setlist{nolistsep}

\newcommand{\netout}{{\widehat{\sigma'}}} 

\usepackage[ruled]{algorithm2e} 

\SetAlFnt{\small}
\SetAlCapFnt{\small}
\SetAlCapNameFnt{\small}
\SetAlCapHSkip{0pt}

\acmJournal{TOG}

\setcopyright{licensedusgovmixed}
\acmYear{2023} \acmVolume{42} \acmNumber{4} \acmArticle{} \acmMonth{8} \acmPrice{15.00}\acmDOI{10.1145/3592141}

\begin{document}
\title{Neural Volumetric Reconstruction for Coherent Synthetic Aperture Sonar}


\author{Albert Reed}
\affiliation{%
 \institution{Arizona State University}
  \country{USA}
}
\email{albertnm123@gmail.com}

\author{Juhyeon Kim}
\affiliation{%
  \institution{Dartmouth College}
  \country{USA}
}
\email{juhyeon.kim.gr@dartmouth.edu}

\author{Thomas Blanford}
\affiliation{%
 \institution{The Pennsylvania State University}
  \country{USA}
}
\email{teb217@psu.edu}

\author{Adithya Pediredla}
\affiliation{%
  \institution{Dartmouth College}
  \country{USA}
}
\email{aditya.eee.nitw@gmail.com}

\author{Daniel C. Brown}
\affiliation{%
  \institution{The Pennsylvania State University}
  \country{USA}
}
\email{dcb19@psu.edu}

\author{Suren Jayasuriya}
\affiliation{%
  \institution{Arizona State University}
  \country{USA}
}
\email{sjayasur@asu.edu}

\begin{abstract}
Synthetic aperture sonar (SAS) measures a scene from multiple views in order to increase the resolution of reconstructed imagery. Image reconstruction methods for SAS coherently combine measurements to focus acoustic energy onto the scene. However, image formation is typically under-constrained due to a limited number of measurements and bandlimited hardware, which limits the capabilities of existing reconstruction methods. To help meet these challenges, we design an analysis-by-synthesis optimization that leverages recent advances in neural rendering to perform coherent SAS imaging. Our optimization enables us to incorporate physics-based constraints and scene priors into the image formation process. We validate our method on simulation and experimental results captured in both air and water. We demonstrate both quantitatively and qualitatively that our method typically produces superior reconstructions than existing approaches. We share code and data for reproducibility.
\end{abstract}



\begin{teaserfigure}
  \includegraphics[width=7in]{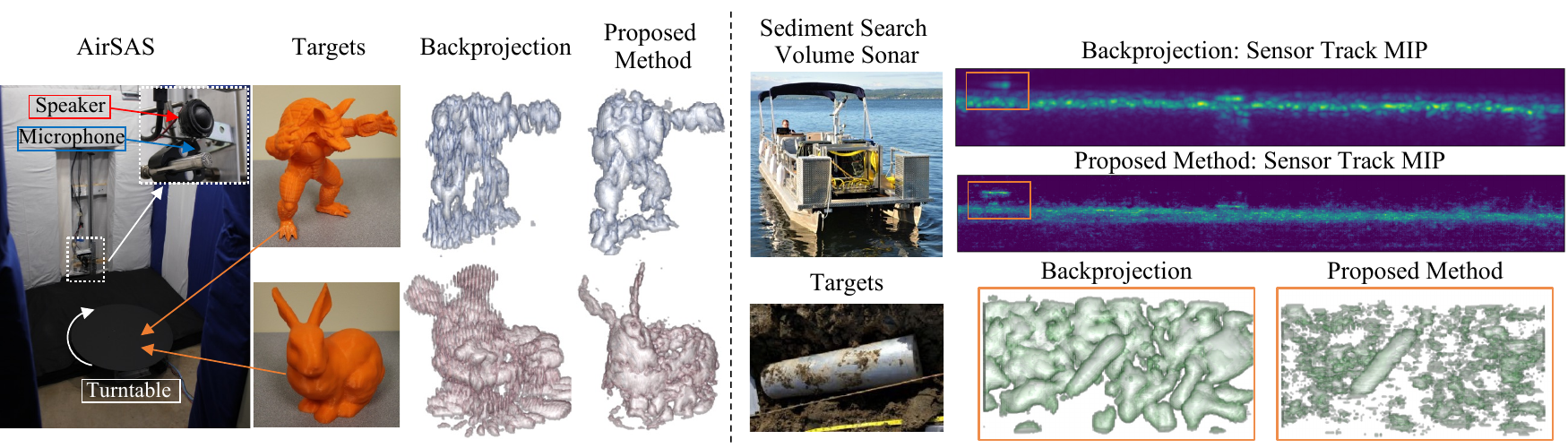}
  \caption{We propose an analysis-by-synthesis optimization that leverages techniques from neural rendering to optimize coherent reconstructions of SAS volumetric scenes. We demonstrate our approach on an in-lab circular SAS in air (AirSAS) and in-water bistatic SAS, the sediment volume search sonar (SVSS). On the left side of the figure, we show the AirSAS, 3D printed targets, and reconstructions obtained using backprojection and our proposed method. On the right side, we show 2D maximum intensity projections (MIPs) of the SVSS track and 3D reconstructions of targets highlighted in orange. In many cases, our method produces better reconstructions than traditional SAS reconstruction algorithms, such as backprojection. SVSS hardware photos courtesy of~\cite{svss-tech-report}.}
  \Description{This is a teaser figure}
  \label{fig:teaser}
\end{teaserfigure}


\maketitle

\section{Introduction}

\begin{figure*}[h!]
  \centering
  \includegraphics[width=7.2in]{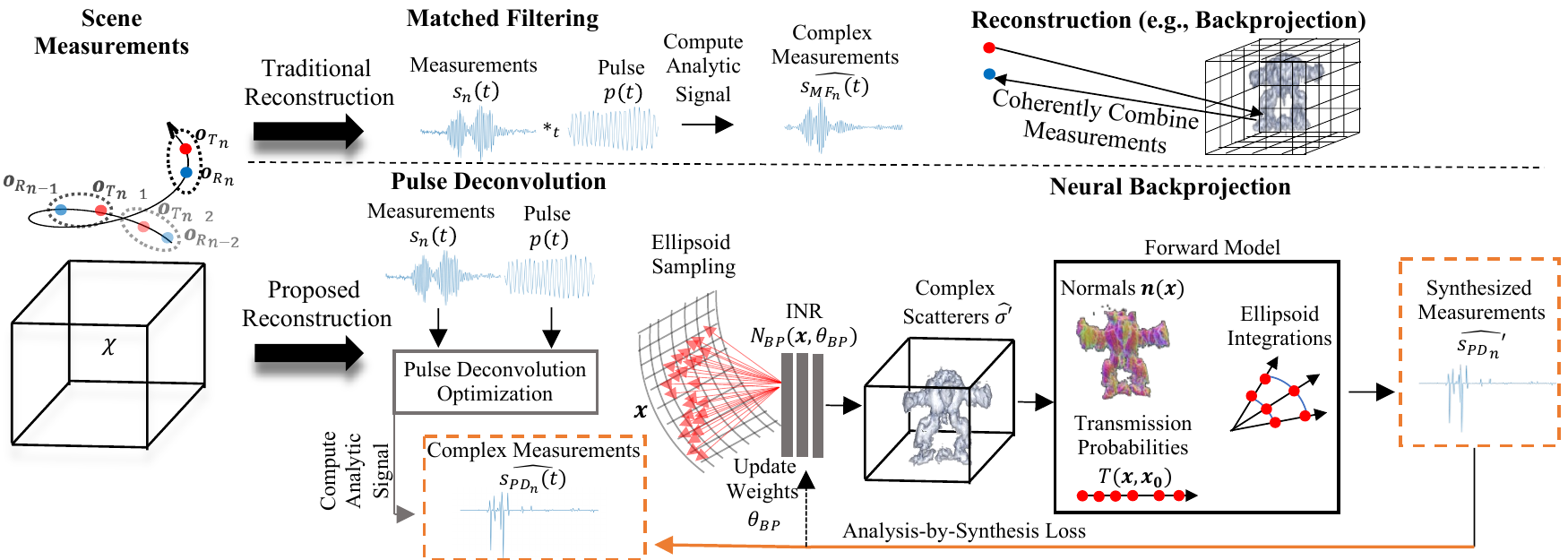}
  \caption{Block diagram of our method. Given measurements obtained from $n$ sensor positions, the top row shows the traditional reconstruction pipeline and the bottom row shows our proposed reconstruction pipeline. Traditional reconstruction uses matched filtering to compress the measurements in time, and then coherently combines measurements using backprojection (or a functionally equivalent algorithm implemented in the Fourier domain). Our method roughly parallels these steps. Instead of matched filtering, we apply pulse deconvolution, which is an optimization that deconvolves the transmitted pulse from measurements. We then propose neural backprojection, which uses a neural network to estimate the scene and synthesizes measurements using our differentiable forward model. The network is trained by minimizing a loss between synthesized and pulse deconvolved measurements.}
  \label{fig:block-diagram}
\end{figure*}

Synthetic aperture sonar (SAS) is an active acoustic imaging technique that coherently combines data from a moving array to form high-resolution imagery, especially of underwater environments~\cite{bellettini2008design, hayes2009synthetic}. The moving array in SAS collects both magnitude and phase information which allows coherent integration methods to achieve resolution parallel to the sensor path that is independent of range~\cite{Callow2003}. These high resolution SAS reconstructions are important for applications in target localization~\cite{williams2016underwater}, monitoring man-made~\cite{nadimi2021efficient} and biological structures~\cite{sture2018recognition}.

However, reconstructing SAS imagery from measurements is an under-constrained problem~\cite{Callow2003}. First, SAS scenes are often undersampled due to the slow propagation of sound relative to the traveling velocity of the sensor platform~\cite{Callow2003, Putney2005}. Second, SAS transducers (transmitter and receiver) are bandlimited, limiting the ability to reconstruct arbitrarily fine spatial frequencies in imagery~\cite{Heering, SINR, Pailhas2017}. 

In related fields, under-constrained imaging problems are typically addressed by optimizing an objective function that incorporates scene or physics-based priors into the reconstruction~\cite{bouman2022foundations}. In particular, a common strategy is analysis-by-synthesis, where a forward model synthesizes measurements from an estimated scene, and the scene is optimized by minimizing the loss between synthesized and given measurements~\cite{bertero2021introduction, lucas2018using}. This framework can flexibly incorporate losses corresponding to sensor noise models and custom scene priors~\cite{kaipio2004computational}. State-of-the-art analysis-by-synthesis typically uses a neural network coupled with a differentiable forward map~\cite{xie2022neural}. Notably, neural radiance fields (NeRF) use a neural network and differentiable volume rendering to synthesize novel views of 3D scenes from 2D images~\cite{mildenhall2021nerf}.

Despite the success of these methods, existing SAS reconstruction algorithms do not use analysis-by-synthesis optimization for reconstruction. Instead, SAS reconstruction coherently combines measurements (in either the time or frequency domain) to focus acoustic energy into the scene. While post-processing optimizations are used (e.g. image-space deconvolution~\cite{SINR, Heering, Pailhas2017}, autofocus~\cite{marston2014autofocusing, Mansour2018, fienup2000synthetic, gerg2020deep}), these methods seek to improve already reconstructed SAS images, rather than incorporate the physics and priors into the image formation.

The challenge for analysis-by-synthesis reconstructions for SAS is the lack of differentiable forward models that are computationally efficient. Acoustic rendering by solving wave equations is physically-realistic but computationally expensive. While geometric acoustic renderers exist for simulating SAS measurements~\cite{gul2017underwater, woods2020ray}, they are not typically differentiable. Further, since SAS measurements can be collected on arbitrary paths~\cite{Callow2003, hayes2009synthetic} (although they typically align with circular or linear scans in practice), it is difficult to precompute a simple forward model. Moreover, SAS arrays transmit spherically propagating sources and measure time series of the received sound pressure. This requires more burdensome sampling than in the optical domain, where cameras measure the intensity of a light ray that travels in a straight line, typically without phase information. These challenges must be taken into account when designing a forward model for analysis-by-synthesis optimization. 

Inspired by recent neural rendering techniques, we design an analysis-by-synthesis method for coherent SAS reconstruction. We leverage implicit neural representations (pioneered by NeRF~\cite{mildenhall2021nerf}) to estimate acoustic scatterers in a 3D volume. Specifically, we formulate a differentiable forward map of a point-based sonar scattering model, and design an ellipsoidal sampling technique to importance sample propagating acoustic pressure waves and then synthesize measurements. Our analysis-by-synthesis optimization allows us to incorporate the physics of the measurement formation, various noise models, and scene priors into the reconstruction. We demonstrate that our technique is not restricted to any particular spatial sampling pattern. 

Our proposed method consists of two primary steps. First, we leverage iterative pulse deconvolution to perform pulse compression which increases the bandwidth of the emitters and sensors computationally. Second, we propose an analysis-by-synthesis reconstruction algorithm, which we term neural backprojection. Our method is inspired by traditional NeRF techniques but varies significantly in terms of sampling (ellipsoidal vs. line sampling) and output (intensity vs. time series). We perform multiple experiments on simulated and hardware-measured data to show both quantitatively and qualitatively that our design outperforms traditional techniques. Using ablation studies, we validate our design and contextualize the performance of our method against existing approaches. Our specific contributions are as follows:
\\
\indent (1) An analysis-by-synthesis framework for incorporating the physics of acoustic signal formation, scene priors, and noise models for coherent volumetric SAS reconstructions. 
\\
\indent (2) Design of a differentiable acoustic forward model that assumes ideal pulse deconvolution to sample along constant time-of-flight ellipsoids.
\\
\indent (3) Evaluation of our proposed method on simulated and real hardware measurements from an in-air circular SAS and in-water measurements from a field survey of a lakebed using a bistatic SAS.
\\
\indent (4) Code and datasets, shared as supplementary material and on the website\footnote{https://awreed.github.io/Neural-Volumetric-Reconstruction-for-Coherent-SAS/}, for reproducibility.


\paragraph{Organization:} The rest of the paper is organized as follows. Section~\ref{sec:Background} provides background on SAS and existing reconstruction methods. Section~\ref{sec:Forward Model} introduces our forward model and provides background on how it is traditionally inverted. Section~\ref{sec:methodoverview} gives a summary of our proposed approach, which involves pulse deconvolution (Section~\ref{sec:pulsedeconvolution}) and neural backprojection (Section~\ref{sec:neuralbackprojection}). Section~\ref{sec:implementation} discusses our datasets and code implementation. Section~\ref{sec:experimentalresults} validates our method by comparing it to existing reconstruction methods and through extensive ablations on hardware measured and simulated data. Section~\ref{sec:discussion} discusses the main findings, impact, and limitations of our work, along with potential future directions. 

\section{Related Work}
\label{sec:Background}
\subsection{Synthetic Aperture Sonar (SAS)}

Many sensing modalities leverage a combination of spatially distributed measurements to enhance performance. In particular, array or aperture processing use either a spatial array of sensors (i.e. real aperture) or a virtual array from moving sensors (i.e. synthetic aperture) to reconstruct a spatial map of the scene. Synthetic aperture imaging has been used for sonar and radar~\cite{gough1997unified}, ultrasound~\cite{jensen2006synthetic} and optical light fields~\cite{levoy1996light}. Synthetic aperture imaging techniques parallel many of those in tomographic imaging which leverage penetrating waves to image the target scene~\cite{ferguson2009generalized}. 

In this paper, we focus on synthetic aperture sonar (SAS) which is a leading technology for underwater imaging and visualization~\cite{hayes2009synthetic}. Acoustical waves are used to insonify the scene, and the time-of-flight of the acoustic signal is used to help determine the location of target objects and scatterers in the scene. Exploiting the full resolution of synthetic aperture systems requires coherent integration of measurements --- combination considering measurement magnitude and phase~\cite{hawkins1996synthetic}. In particular, coherent integration yields an along-track (along the sensor path) resolution independent of range and wavelength~\cite{hawkins1996synthetic}. 

SAS systems typically transmit \textit{pulse compressable} waveforms, waveforms with large average power but with good range resolution~\cite{eaves2012principles, harrison2019introduction}. Common examples include swept frequency waveforms, which apply a linear or non-linear change in waveform frequency over time~\cite{harrison2019introduction}. These waveforms are pulse compressed at the receiver by correlating measurements with the transmitted waveform (i.e., pulse). This processing is commonly referred to throughout communications and remote sensing communities as matched filtering (or replica-correlation). Waveform design is an active areas of research for creating optimal compressed waveforms --- there is a tradeoff between range resolution and hardware limitations affecting bandwidth~\cite{Callow2003}.

\subsection{SAS Reconstruction}
\label{sec:sas-reconstruction}
Many algorithms exist for reconstructing imagery from SAS measurements. Perhaps the most intuitive algorithm is time-domain backprojection (also called delay-and-sum or time-domain correlation) which backprojects received measurements to image voxels using their time-of-flight measurements. The advantage of this method is that it works for arbitrary scanning geometries, however, traditionally it has been considered slow to compute~\cite{hayes2009synthetic, soumekh1999synthetic}. Wavenumber domain algorithms such as range-Doppler and $\omega-k$ migration are significantly faster but require assuming a far-field geometry and an interpolation step to snap measurements onto a Fourier grid~\cite{eaves2012principles, hayes2009synthetic}. For circular scanning geometries (CSAS), specialized reconstruction algorithms~\cite{plotnick2014fast, marston2011coherent, marston2014autofocusing} exploit symmetry and connections to computed tomography~\cite{ferguson2005application} for high-resolution visualization. Even further specialized SAS techniques leverage interferometry~\cite{hansen2003signal,griffiths1997interferometric} for advanced seafloor mapping. 
In this paper, we use time-domain backprojection as our baseline SAS reconstruction approach. While this method is considered slow conventionally, modern computing capabilities with GPUs have alleviated this bottleneck~\cite{gerg2020gpu}. Backprojection is applicable to nearly arbitrary measurement patterns, in contrast with Fourier-based methods which make a collocated transmit/receive assumption and require interpolation to a Fourier grid. Additionally, backprojection and Fourier methods typically produce equivalent imagery for data that meets the requirements necessary of the Fourier-based algorithms~\cite{bamler1992comparison}. 

Many methods exist for further improving the visual quality of reconstructed imagery. Notably, many methods estimate the platform track and motion to correct imaging errors~\cite{cook2007synthetic, brown2019interpolation, yu2006multiple, cook2008analysis, fienup2000synthetic, Callow2003, fortune2001statistical},deconvolution~\cite{Putney2005,Heering}, autofocus~\cite{marston2014autofocusing,gerg2021real}, and accounting for environment noise~\cite{chaillan2007speckle, Callow2003, hayes1992broad, piper2002detection}. These methods are complementary to our reconstruction approach, and could be investigated further in conjunction with our pipeline.

\subsection{Acoustic Rendering}
\label{sec:acousticrendering}
Modeling acoustic information in an environment has largely fallen into two broad categories: geometric acoustics and wave-based simulation. Geometric acoustic methods, also known as ray tracing, are based upon a small wavelength approximation to the wave equation~\cite{savioja2015overview,liu2020sound}. The analog of Kajiya's rendering equation for room acoustics has been proposed with acoustic bidirectional reflectance distributions~\cite{siltanen2007room}. Further, bidirectional path tracing has been introduced to handle occlusion in complex environments~\cite{cao2016interactive}. However, diffraction can cause changes in sound propogation, particularly near edges where sound paths bend. To account for this, researchers have proposed techniques to add these higher-order diffraction effects to path tracing and radiosity simulations~\cite{schissler2014}. 

In contrast, solving the wave equation directly encapsulates all these diffraction effects, but is computationally expensive~\cite{hamilton2017}. To alleviate processing times, precomputation has been used extensively~\cite{raghu2010,raghu2014} with these systems. In addition, acoustic textures have been introduced to enable fast computation of wave effects for ambient sound and extended sources~\cite{zhang2018ambient,zhang2019acoustic}. Further, anisotropic effects for complex directional sources can be rendered efficiently~\cite{chaitanya2020}. In addition to acoustically modeling large environments, there has been a large body of work modeling the vibration modes of complex objects~\cite{wang2018toward}. This includes elastic rods~\cite{schweickart2017animating}, fire~\cite{chadwick2011animating}, fractures~\cite{zheng2010rigid}, thin shells~\cite{chadwick2009harmonic}. 

For SAS in particular, there have been several acoustic rendering models proposed in the literature. The Personal Computer Shallow Water Acoustic Tool-set (PC-SWAT) is a common underwater simulation environment that leverages finite element modeling~\cite{pcswat} as well as extensions to ray-based geometric acoustics~\cite{woods2020ray}. HoloOcean is a more general underwater robotics simulator that enables simulation of acoustics~\cite{potokar2022holoocean}. BELLHOP is a popular acoustic ray tracing model for long range propagation modeling~\cite{gul2017underwater}. In this work, we leverage the Point-based Sonar Scattering Model (PoSSM), a single bounce acoustic scattering model~\cite{brown2017point, brown2017modeling, johnson2018sas}, to design our forward model for our neural backprojection method.

\subsection{Transient and Non-Line-of-Sight Imaging}
Many works use optical transient imaging for measuring scenes in depth by leveraging continuous wave time-of-flight devices~\cite{heide2013low, kadambi2016occluded, kadambi2013coded} or impulse-based time-of-flight single photon avalanche diode (SPADs)~\cite{o2017reconstructing, callenberg2021low}. In particular, transient imaging is useful for non-line-of-sight (NLOS) reconstruction~\cite{velten2012recovering, arellano2017fast, lindell2019wave, ahn2019convolutional, liu2019non, pediredla2019ellipsoidal}. Analysis-by-synthesis optimization has been effective for NLOS problems including differentiable transient rendering~\cite{iseringhausen2020non, yi2021differentiable, plack2023fast,wu2021differentiable} and even utilized for conventional cameras~\cite{sreenithy-bmvc, chen2019steady}. While there are interesting connections between transient/NLOS imaging and SAS, more research is needed to connect these domains. Lindell et al. used acoustic time-of-flight to perform NLOS reconstruction~\cite{lindell2019acoustic}, but do not consider SAS processing. SAS imaging presents new technical challenges for transient imaging including non-linear measurement trajectories and bi-static transducer arrays, coherent processing, and acoustic-specific effects.

\subsection{Neural Fields}
\label{sec:INRbackground}
Recently, there has been large interest in representing scenes or physical quantities using the optimized weights of neural networks~\cite{xie2022neural}. These networks, termed implicit neural representations (INR), or more broadly as neural rendering or neural fields, exploded in popularity following NeRF, which used them for learning 3D volumes from 2D images~\cite{mildenhall2021nerf}. These networks use a positional encoding to overcome spectral bias~\cite{cao2019towards}. 


INRs have been used in a huge number of inverse problems across imaging and scientific applications~\cite{xie2022neural}. In particular, INRs have been recently applied to tomographic imaging methods which has similarities to synthetic aperture processing~\cite{sun2022coil,zang2021intratomo,ruckert2022neat,reed4Dct}. Of particular interest to our work is neural rendering for time-of-flight (ToF) cameras. Time-of-Flight Radiance Fields couples a ToF camera with an optical camera to create depth reconstructions~\cite{attal2021torf}. While this work does consider the phase of the ToF measurements, their method does not feature coherent integration of phase values as we do in synthetic aperture processing. Further, they consider ToF cameras where each pixel corresponds to samples along a ray. In contrast, measurements from a SAS array correspond to samples along propagating spherical wavefronts. Shen et al. propose leveraging neural fields for NLOS imaging~\cite{shen2021nonlineofsight}. In contrast, we leverage neural fields coupled with differentiable acoustic forward model for SAS imaging.

Several works consider apply neural fields for sonar and SAS image reconstruction. Reed et al. leverage neural fields to perform 2D CSAS deconvolution~\cite{sinrconference,SINR}. Their method post-processes (deblurs) reconstructed 2D scenes for circular SAS measurement geometries~\cite{SINR}. On the other hand, our proposed approach focuses on reconstruction for 3D SAS. Recently, a method using an INR for forward-looking sonar was developed~\cite{qadri2022neural}. Forward-looking sonar stitches images together from individual slices, and does not typically utilize coherent integration. Our method differs from this method as we account for the effects of the transmit waveform and consider the coherent integration of multiple views, which is fundamental to synthetic aperture processing. 



\section{Background: Forward Model and Time-Domain Backprojection}
\label{sec:Forward Model}

\begin{table}[]
\caption{Definition of operators and variables used throughout the paper.}
\begin{tabular}{ll}
\hline
\textbf{Operators}   & \textbf{Definition}                                                                                      \\ \hline
$\widehat{x}$        & Complex-valued analytic signal of $x$.                                                                   \\
|x|                  & Magnitude of $x$. If $x = a + jb$ then $|x| = \sqrt{a^2 + b^2}$.                                         \\
$\angle x$           & Angle of $x$. If $x = a + jb$ then $\angle x = \tan^{-1}{b/a}$.                                          \\
\textbf{||x||}       & 2-norm of vector $\mathbf{x}$.                                                                           \\
$\mathcal{H}(x)$     & Hilbert transform of $x$.                                                                                \\
$\mathcal{R}(x)$     & Real part of $x$.                                                                                        \\
$\mathcal{I}(x)$     & Imaginary part of $x$.                                                                                   \\ \hline
\textbf{Variables}   &                                                                                                          \\ \hline
$p(t)$               & Transmitted pulse.                                                                                       \\
$s_n(t)$             & Raw measurements of sensor $n$.                                                                          \\
$s_{{\text{MF}}}(t)$ & Match-filtered measurements.                                                                             \\
$s_{\text{PD}}(t)$   & Given Pulse deconvolved measurements .                                                                         \\
$\mathbf{s'}_{\text{PD}}(t)$   &  Synthesized pulse deconvolved measurements.                                                                        \\
$N_{\text{PD}}$      & Pulse deconvolution network.                                                                             \\
$N_{\text{BP}}$      & Neural backprojection network.                                                                           \\
$\Delta f$           & Transmit pulse bandwidth.                                                                                \\
$f_\text{start}$     & Transmit pulse start frequency.                                                                          \\
$f_\text{stop}$      & Transmit pulse stop frequency.                                                                           \\
$f_c$                & Transmit pulse center frequency, $(f_\text{start} + f_\text{stop})/2$.                                   \\
$\mathcal{X}$        & The set of all scene points.                                                                             \\
$\mathbf{E}_{r_i}$   & \begin{tabular}[c]{@{}l@{}}The set of points $x$ on the ellipsoid surface\\ at range $r_i$.\end{tabular} \\

$b_T(\mathbf{x})$    &  Transmitter directivity function at point $\mathbf{x}.$
        \\

$b_R(\mathbf{x})$   &   Receiver directivity function at point $\mathbf{x}.$
        \\

$a(r_i)$             & Length of ellipsoid $x$ semi-axis defined at range $r_i$.                                                \\
$b(r_i)$             & Length of ellipsoid $y$ semi-axis defined at range $r_i$.                                                \\
$c(r_i)$             & Length of ellipsoid $z$ semi-axis defined at range $r_i$.                                                \\
$\mathbf{o}_T$       & Transmitter origin.                                                                                      \\
$\mathbf{x}_T$       & Transmission ray.                                                                                        \\
$\mathbf{o}_R$       & Receiver origin.                                                                                         \\
$\mathbf{x}_R$       & Receive ray.                                                                                             \\
$\mathbf{d}_T$       & Transmit ray direction (unit vector).                                                                    \\
$T(\mathbf{o}_T, \mathbf{x})$            &   \begin{tabular}[c]{@{}l@{}}Transmission probability from a \\ transmitter origin to a point $\mathbf{x}$.\end{tabular}
        \\
$T(\mathbf{o}_R, \mathbf{x})$              &   \begin{tabular}[c]{@{}l@{}}Transmission probability from a \\ receiver origin to a point $\mathbf{x}$.\end{tabular}
        \\
$r_i$                & Distance $i$ from transmit and receiver origin.                                                          \\
$l_i$                & Depth $i$ along ray.                                                                                     \\
$\widehat{\sigma'}$  & Estimated complex scattering function.                                                                   \\
$\sigma$             & Ground truth scattering function.                                                             \\ \hline
\end{tabular}
\label{fig:notation}
\end{table}

We first introduce the forward measurement model that we use later to design our analysis-by-synthesis optimization. This model is inspired by a point-based sonar scattering model~\cite{brown2017modeling,brown2017point}. Point scattering models assume high-frequency propagation (i.e., geometric acoustics), but offer computational tractability and differentiability that is friendly for neural rendering techniques. 


We now formulate our imaging model mathematically (we refer the reader to Table~\ref{fig:notation} for reference to the notation used throughout the text). Let $\mathbf{x} \in \mathbb{R}^{3}$ describe a 3D coordinate in a scene, $\sigma(\mathbf{x}) \in \mathbb{R}$ the amplitude of the acoustic scatterer at $\mathbf{x}$, $p(t)$ is the transmitted pulse, and $\mathcal{X}$ the set of all coordinates in the volume of interest. We also define $b_T(\mathbf{x})$ and $b_R(\mathbf{x})$ to be the transmitter and receiver directivity functions respectively. We define $T(\mathbf{o}_T, \mathbf{x})$ and $T(\mathbf{o}_R, \mathbf{x})$ as the transmission probabilities between a scene point and the transmitter and receiver origins, respectively, where $T(\cdot)$ is a function that computes the transmission probability between two points and enables our model to account for occlusion.

Let $R_T = ||\mathbf{o}_T-\mathbf{x}||$ and $R_R = ||\mathbf{o}_R-\mathbf{x}||$ be the distances between the scene point and sensor transmit and receive origins, respectively. Then, the receiver records real-valued measurements similar to Brown et al.~\cite{brown2017point}:

\begin{equation}
\label{eq:forwardmodel}
    \mathbf{s}(t) = \int\displaylimits_{\mathcal{X}} \frac{b_T(\mathbf{x})b_R(\mathbf{x})T(\mathbf{o}_T, \mathbf{x})T(\mathbf{o}_R, \mathbf{x})}{2\pi R_T R_R}L(\sigma(\mathbf{x}))p\left(t - \frac{R_t + R_R}{c}\right)\mathrm{d}\mathbf{x},
\end{equation}
where $L(\cdot)$ is a Lambertian acoustic scattering model computed using the normal vector at a point $\mathbf{n}(\mathbf{x})$. Contrary to acoustic radiance~\cite{siltanen2007room}, this equation models acoustic pressure which has a $1/\text{distance}$ falloff due to spherical propagation~\cite{pierce1981}. Additionally, note that the sensor measurement $s(t) = s(\mathbf{o}_T,\mathbf{o}_R, t)$ is actually a function of the transmit and receive origins as well. Throughout the text, we will sometimes index measurements as $\mathbf{s}_n(t)$, but typically omit $n$ for brevity.



\begin{figure}
  \centering
  \includegraphics[width=2.6in]{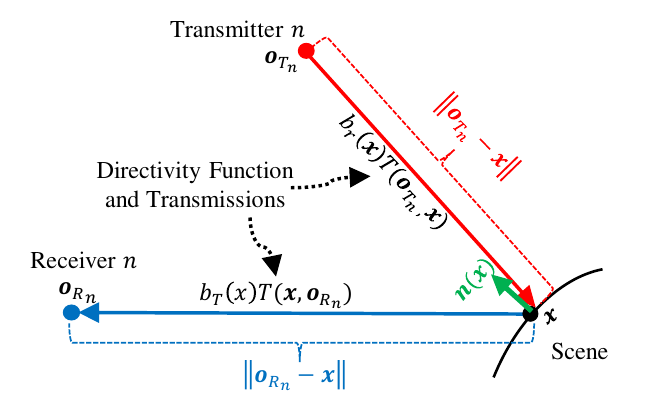}
  \caption{Forward model geometry. A transmitter pings the scene with the transmit waveform. The propagation of the waveform to a scene point $\mathbf{x}$ is weighted by the directivity function and transmission probability. The scene point reflects acoustic energy back towards the receiver that is weighted by a Lambertian scattering term, the return transmission probability, and the receiver directivity.}
  \label{fig:ray-geometry}
\end{figure}


\subsection{Conventional SAS Reconstruction with Time-domain Backprojection}
\label{sec:pulsecompression}
We now briefly discuss the traditional processing pipeline for SAS measurements, where received measurements are compressed in range and the coherent integration of measurements forms an image. 

\paragraph{Match-filtering and Pulse Compression.} The first processing step for the received signal is to perform matched filtering by cross-correlating with the transmitted pulse:
\begin{equation}
    \mathbf{s}_{\text{MF}}(t) = \mathbf{s}(t) \ast_t p^{*}(-t).
\end{equation}
Note that we have written the cross-correlation as a convolution in time $\ast_t$ with the time-reversed conjugate signal $p^{*}(-t)$  which is typically done in sonar/radar processing literature. Match-filtering is a robust method for deconvolving the transmission waveform from measurements~\cite{soumekh1999synthetic} and is the optimal linear filter for detecting a signal in white noise~\cite{smith1997scientist}.

For a simple rectangular transmitted pulse $p(t) = A, -\tau/2 \leq t \leq \tau/2$ and zero elsewhere, it is easy to show that $S_{MF}(t)$ is a triangle function and the energy of the signal is $A^2\tau$. Since the transmitter is operating at peak amplitude (in this example), the duration of the signal $\tau$ yields higher energy, and thus higher signal-to-noise ratio (SNR). However, increasing $\tau$ comes at the expense of poor range-resolution, given by the equation~\cite{eaves2012principles, harrison2019introduction}: \begin{equation}
    \Delta r = \frac{c}{2B}\label{eq:range-resolution}, 
\end{equation}
where $c$ is the pulse propagation speed, and the bandwidth $B = 1/\tau$ in this case. 

To decouple the relationship between range resolution and energy of the signal, sonars transmit a frequency-modulated pulse~\cite{harrison2019introduction}. In particular, the linear frequency modulated (LFM) pulse is a common choice:
\begin{equation}
    p(t) = w(t)\cos\left(2\pi \frac{\Delta f}{2 \tau}t^2 + 2\pi f_\text{start} t\right),
\end{equation}
where the bandwidth in Hz is given by $\Delta f$ = $|f_\text{start} - f_\text{stop}|$, $\tau$ is the pulse duration in seconds, and $w(t)$ is a windowing function to attenuate side-lobes in the ambiguity function. The range-resolution of a pulse-compressed waveform computed using Eq.~\eqref{eq:range-resolution} is $\Delta r = \frac{c}{2 \Delta f}$.

\paragraph{Coherent Backprojection.} Synthetic aperture imaging reconstructs images with range-independent along-track resolution through coherent integration of measurements~\cite{Callow2003}. As the transmitted waveform is typically modulated by a carrier frequency, it is desirable to coherently integrate the envelope of received measurements. The signal envelope can be estimated with range binning~\cite{hayes2009synthetic}, but the analytical form of the envelope is obtained with the Hilbert transform~\cite{bracewell1986fourier}. In particular, the Hilbert transform can be used to obtain the analytic signal (also called the pre-envelope):
\begin{equation}
    \widehat{\mathbf{s}_{\text{MF}}} = \mathbf{s}_{\text{MF}} + j\mathcal{H}(\mathbf{s}_{\text{MF}})
\end{equation}
where $j = \sqrt{-1}$ and $\mathcal{H}$ is the Hilbert transform operator.


Given these (now) complex measurements, SAS image formation uses the sensor and scene geometry to coherently integrate measurements that are projected onto the scene using their time-of-flights,
\begin{equation}
    I_{\text{BP}}(\mathbf{x}) = \sum_n \widehat{\mathbf{s}_{\text{MF}_n}}\left(t - \frac{R_T+R_R}{c}\right). \label{eq:bp}
\end{equation}
Later, we show that this equation effectively integrates energy along ellipsoids defined by the transmit and receive locations and time-of-flights. We remind the reader that the values $R_T,R_R$ are defined in terms of $\mathbf{x}$ and the transmit and receive positions of the transducers, and thus are not constant for differing $n$ and 
$\mathbf{x}$. Eq.~\eqref{eq:bp} is the coherent integration of measurements and results in a complex image. The final estimate of the acoustic scattering coefficient $\sigma(\mathbf{x})$ is obtained by taking the image magnitude $|I_{\text{BP}}(\mathbf{x})|$~\cite{hawkins1996synthetic}. 



\section{Overview of Proposed Method}
\label{sec:methodoverview}

Our proposed reconstruction method (shown in Fig.~\ref{fig:block-diagram}) consists of two main steps that roughly parallel the matched filtering and coherent backprojection described in the previous section. First, we propose deconvolving given waveforms via an iterative deconvolution optimization rather than performing matched filtering. While matched filtering can be optimized through waveform design to realize a better ambiguity function in cross-correlation (i.e. better range compression), these techniques require a priori knowledge and do not work across a variety of sonar environments. In contrast, we present an adaptable approach to waveform compression where performance can be tuned via sparsity and smoothness priors, which we label pulse deconvolution in Section~\ref{sec:pulsedeconvolution}.

Our second step is an analysis-by-synthesis reconstruction using an implicit neural representation (similar to NeRF in traditional view synthesis~\cite{mildenhall2021nerf}). We use a network to predict complex-valued scatterers, and use a differentiable forward model to synthesize complex sensor measurements in time. Traditional NeRF scene sampling methods are not directly applicable to our problem since we require sampling the scene points with constant time-of-flight, which correspond to ellipsoids with the transmitter and receiver as foci~\cite{pediredla2019ellipsoidal}. Thus, we project rays from the transducer and sample rays at the intersection of ellipsoidal surfaces corresponding to measurement time bins and develop importance sampling methods to determine transmission probabilities for these rays and ellipsoidal surfaces. Finally, we explain how we implement our physics-based priors, such as a Lambertian scattering assumption, and regularization to our analysis-by-synthesis optimization.

\begin{figure}
  \centering
  \includegraphics[width=3.45in]{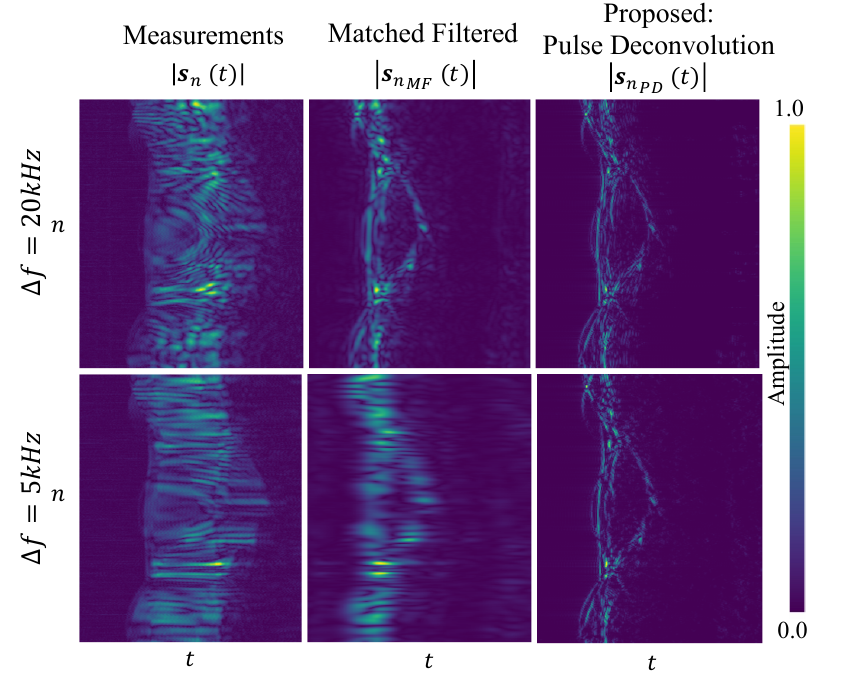}
  \caption{Comparing matched filtering performance with pulse deconvolution. We show the magnitude (i.e., signal envelope) of AirSAS measurements of a 3D-printed Stanford bunny measured with an LFM ($f_c=20$~kHz) at bandwidths $\Delta f = 20$~kHz (top row) and $\Delta f = 5$~kHz (bottom row). Matched filtering's ability to compress measurements degrades at the lower $5$~kHz bandwidth. Our pulse deconvolution method yields better compression at both bandwidths.}
  \label{fig:deconv-vs-mf}
\end{figure}

\section{Pulse Deconvolution}
\label{sec:pulsedeconvolution}

We now present our method for deconvolving the transmit waveform from SAS measurements. We refer to the deconvolved measurements as $\mathbf{s}_{\text{PD}}(t)$. We propose optimizing a network, labeled the pulse deconvolution network $N_{pd}$ as,
\begin{align}
\begin{split}
    \mathcal{L}_{PD} = ||N_{PD}(t;\theta_{\text{PD}}) \ast p^{*}(-t) - \mathbf{s}(t)||_2 \\
    + \lambda_{1_{\text{PD}}}\mathcal{L}_{\text{Sparse}}^{\text{PD}} + \lambda_{2_{\text{PD}}}\mathcal{L}_{\text{TV Phase}}^{\text{PD}}, \label{eq:range-deconvolution}
\end{split}
\end{align}
where $\theta_{\text{PD}}$ are the trainable weights of the network. The sparsity and phase smoothing operators are defined as,
\begin{align}
    \mathcal{L}_{\text{Sparse}}^{\text{PD}} = \sum_t||N_{PD}(t;\theta_{\text{PD}})||_1 \\
    \mathcal{L}_{\text{PD, TV Phase}}^{\text{PD}} = \sum_t||\nabla \angle N_{PD}(t;\theta_{\text{PD}})||_1,
    \label{eq:a-by-s-deconv}
\end{align} 
where $\angle \cdot$ denotes the angle of a complex value
and where regularizors are weighted by scalar hyperparameters $\lambda_{1_{\text{PD}}}$ and $\lambda_{2_{\text{PD}}}$. We find that sparsity regularization is particularly important for recovering accurate deconvolutions. The total-variation (TV) phase prior encourages the solution to have a smoothly varying phase, which we find helps attenuate noise in the deconvolved waveforms. We minimize the total pulse deconvolution loss $\mathcal{L}_{PD}$ with respect to the network weights $\theta_{\text{PD}}$ using PyTorch's ADAM~\cite{kingma2014adam} optimizer. 

We implement the network using an implicit neural representation (INR), where input coordinates are transformed with a hash-encoding~\cite{mueller2022instant} to help the network overcome spectral bias. Our implementation trains an INR per batch of sensor measurements. In particular, we add an additional input to the network, $n$ (omitted from Eq~\eqref{eq:a-by-s-deconv}), that denotes a sensor index and allows a single network to deconvolve a batch of sensor measurements. At inference, we form the pulse deconvolved signal $\mathbf{s}_{\text{PD}}(t) = N_{PD}(t;\theta_{\text{PD}})$ using the network and then calculate the analytic signal $\widehat{\mathbf{s}_{\text{PD}}}$ to be used for coherent neural backprojection described in Section~\ref{sec:neuralbackprojection}. 

Fig.~\ref{fig:deconv-vs-mf} compares the pulse compression performance of match-filtering and our deconvolution method on AirSAS (Section~\ref{subsec:airsas}) data. The figure shows $n=360$ bunny measurements recorded using an LFM with center frequency $f_c=20$ kHz at bandwidths $\Delta f = 20$ kHz and $\Delta f = $ 5 kHz. In agreement with Eq.~\eqref{eq:range-resolution}, match-filtering compresses measurements more at $\Delta f = 20$ kHz than at $\Delta f = 5$ kHz. Our proposed deconvolution method compresses the measurements more than match-filtering and with remarkably similar performance at both bandwidths.


In addition to our proposed deconvolution method, we experiment with deconvolving the waveforms in a single step using simple Wiener deconvolution, but observe notably inferior performance compared to the network. We also tried optimizing Eq.~\eqref{eq:range-deconvolution} without a neural network (i.e., by updating the values at each time bins directly via gradient descent) and actually observed competitive deconvolution performance. However, we find that the network seems to output marginally smoother deconvolved waveforms. Given this observation, and the fact that the INR had a nearly equivalent latency, we use a network in Eq.~\eqref{eq:range-deconvolution} to obtain $\mathbf{s}_{\text{PD}}(t)$ for all experiments. 





\section{Neural Backprojection}
\label{sec:neuralbackprojection}





Our first choice for synthesizing measurements is to use the point-based scattering model of Eq.~\eqref{eq:forwardmodel}. However, coherent backprojection methods integrate the envelope of the signal from Eq.~\eqref{eq:forwardmodel}, and thus we perform our analysis-by-synthesis optimization by computing a loss between synthesized measurements and given analytic (i.e. complex) signal measurements from the data. To synthesize the analytic measurements, we derive an approximation to the analytic forward model (see the supplemental material for the full derivation): 
\begin{align}
   \widehat{\mathbf{s'}_{\text{PD}}}\left(t=\frac{R_T+R_R}{c}\right) \approx \int_{\mathbf{E}_{r}} b_T(\mathbf{x})b_R(\mathbf{x})T(\mathbf{o}_T, \mathbf{x})T(\mathbf{x}, \mathbf{o}_R)L(\netout(\mathbf{x}))\mathrm{d}\mathbf{x},
   \label{eq:data-consistency}
\end{align}
where $\widehat{\mathbf{s'}_{\text{PD}}}$ are synthesized (i.e. rendered) complex-valued pulse deconvolved measurements. This equation synthesizes measurements using the transmitter and receiver directivity functions $b_T(\mathbf{x})$ and $b_R(\mathbf{x})$, the transmission probabilities to and from each point $T(\mathbf{o}_T, \mathbf{x})$ and $T(\mathbf{x}, \mathbf{o}_R)$, and a Lambertian scattering function $L(\cdot)$ applied to complex-valued scene scatterers $\netout$.

A key property of Eq.~\eqref{eq:data-consistency} is that $\netout$ now defines complex-valued scatterers, which is consistent with how conventional backprojection algorithms recover a complex SAS image corresponding to the complex envelope of the matched filtered signals~\cite{bracewell1986fourier, hayes1992broad}. It also accounts for any non-idealities in the pulse deconvolution which can introduce complex magnitude and phase into the equation. Thus in Eq.~\eqref{eq:data-consistency}, we are synthesizing complex-valued estimates of given deconvolved measurements such that $\mathcal{R}(\widehat{\mathbf{s'}_{\text{PD}}})(t) \approx \mathbf{s}_{\text{PD}}(t)$. This enables us to coherently integrate scatterers and recover our estimate of the scatterers $\sigma(x)$ from Eq.~\eqref{eq:forwardmodel} by computing the magnitude $|\netout(\mathbf{x})| \approx \sigma(\mathbf{x})$. 

In Eq.~\ref{eq:data-consistency}, the measured amplitude at a particular time is given by integrating complex-valued scatterers along a 3D ellipsoid surface $\mathbf{E}_{r}$. Assuming no multipath, the ellipsoid is defined by a constant time-of-flight from the transmitter and receiver origins (and whose sampling we detail further in Section~\ref{sec:ellipsoid}). The ellipsoid approximation assumes that pulse deconvolution works well, and we show in our experimental results that not performing pulse deconvolution results in worse reconstructions. Finally, we note that we assume $b_R(\mathbf{x})$ = 1 for all $\mathcal{X}$, which is reasonable since receivers typically have relatively large beamwidths to suppress aliasing~\cite{gough1997imaging}, and omit the term $\frac{1}{2\pi R_T R_R}$ in the equation as this is commonly done in time-domain beamformers in actual implementation~\cite{hayes1992broad}.


We estimate the complex scattering function $\netout$ using a neural network, entitled the back-projection network $N_{\text{BP}}$, that is parameterized with weights $\theta_{\text{BP}}$. Specifically, the network defines the complex scatterer at each location,
\begin{equation}
    \netout(\mathbf{x}) = N_{\text{BP}}(\mathbf{x}; \theta_{\text{BP}}).
\end{equation}
Thus the analysis-by-synthesis optimization loss can be written as 
\begin{equation}
    \mathcal{L}_{BP} = \left|\left|\widehat{\mathbf{s'}_{\text{PD}}} - \widehat{\mathbf{s}_{\text{PD}}} \right|\right|_2,
    \label{}
\end{equation}
where we minimize the loss between complex-valued synthesized and given pulse deconvolved measurements with respect to the network weights $\theta_{\text{BP}}$ using PyTorch's ADAM~\cite{kingma2014adam} optimizer. 

In the next subsections, we describe how we importance sample the scene via ellipsoids of constant time-of-flight (Section~\ref{sec:ellipsoid}), estimate the transmission probabilities for the transmit and return rays and compute surface normals (Section~\ref{sec:occ-normal}), and compute the above loss with regularization terms (Section~\ref{sec:regularization}).

\begin{figure}
  \centering
  \includegraphics[width=2.9in]{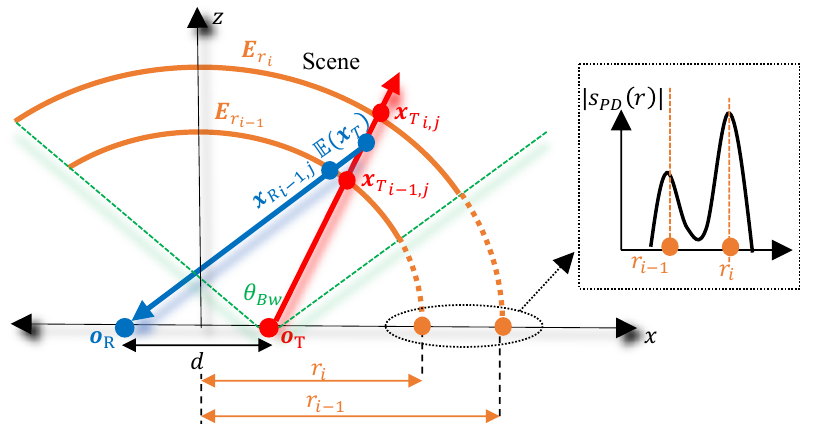}
  \caption{2D diagram ($xz$ slice) of our sampling scheme. A transmission ray (red) is projected within the transmitter beamwidth (green) and sampled at the intersection of the ellipsoids (orange) defined by range samples (shown in the dotted black box). The return ray (blue) is computed from the expected depth of the transmit ray back to the receiver.}
  \label{fig:sampling-scheme}
\end{figure}

\subsection{Ellipsoidal Sampling}
\label{sec:ellipsoid}
As shown in Fig.~\ref{fig:sampling-scheme}, a set of points for a constant time-of-flight $t=\frac{R_t + R_R}{c}$ define an ellipsoid whose foci are the transmitter and the receiver positions, and with semi-major axis length $r = c\cdot t/2$, where $c$ is the sound speed. The ellipsoid can be written as:
\begin{equation}
    \frac{x^2}{a(r)^2} + \frac{y^2}{b(r)^2} + \frac{z^2}{c(r)^2} - 1 = 0,
    \label{eq:ellipse-equation}
\end{equation}
where transmit $\mathbf{o}_T$ and receive $\mathbf{o}_R$ elements are separated by distance $d$. The ellipsoid semi-axes lengths are, 
\begin{equation}
    a(r) = r, b(r)=\sqrt{(r)^2 - (d/2)^2}, c(r) = b(r).
\end{equation}

Thus, our problem reduces to sampling the intersection of transmitted and received rays with these ellipsoids. We begin by sampling a bundle of rays originating from the transmitter and within its beamwidth $\theta_{\text{bw}}$. Fig.~\ref{fig:sampling-scheme} illustrates a transmitted ray in red with direction $\mathbf{d}_{{T}_j}$ and defined as 
\begin{equation}
    \mathbf{x}_{T_{ij}} = \mathbf{o}_T + l_i \mathbf{d}_{{T}_j}
    \label{eq:tx-ray}
\end{equation}
where $l_i$ are depth samples along the ray. We sample this ray at its intersection with ellipsoids defined by desired range of samples. Note that we index ray samples by the ray direction $j$ and the depth sample $i$.

In contrast to conventional NeRF methods that use a coarse network for depth importance sampling, we can use the time series measurements. In particular, we sample time $t_i$ with probability $Pr(t = t_i) = \frac{|\widehat{\mathbf{s}_{\text{PD}}}(t_i)|}{\sum_i\left|\widehat{\mathbf{s}_{\text{PD}}}(t_i)\right|}$. 

This concept is illustrated in Fig.~\ref{fig:sampling-scheme}. In the upper left of the figure, we show a deconvolved waveform that is sampled at two-time bins (converted to range using $r_i = c\cdot t_i/2$). These sampled ranges define ellipsoids drawn in the orange curves. We find the depth $l_i$ that a ray intersects the ellipsoid by substituting the ray into the ellipsoid equation. Substitution yields a quadratic equation with a positive root,
\begin{equation}
    l_i = -b_0 + \frac{\sqrt{b_0^2 - 4a_0c_0}}{2a_0}
    \label{eq:quadratic}
\end{equation}
where 
\begin{gather}
    a_0 = \frac{[d_T]_x^2}{a(r_i)^2} + \frac{[d_T]_y^2}{b(r_i)^2} + \frac{[d_T]_z^2}{c(r_i)^2} \\
    b_0 = \frac{2[x_T]_x [d_T]_x}{a(r_i)^2} + \frac{2[x_T]_y [d_T]_y}{b(r_i)^2} + \frac{2[x_T]_z [d_T]_z}{c(r_i)^2} \\
    c_0 = \frac{[x_T]_x^2}{a(r_i)^2} + \frac{[x_T]_y^2}{b(r_i)^2} + \frac{[x_T]_y^2}{c(r_i)^2} - 1,
\end{gather}
and the notation $[d_T]_x$ refers to the $x$ component of the vector $\mathbf{d}_{T}$. The positive root of the quadratic corresponds to the valid intersection, while the negative root is the intersection on the other side of the ellipsoid.


While not shown in the figure, we also implement a simple direction-based priority sampling. Specifically, we sample a set of sparse rays spanning uniform directions within the transmitter beamwidth. We integrate along each ray and use the resulting magnitude to weight the likelihood of dense sampling in nearby directions. 



\begin{figure}
  \centering
  \includegraphics[width=2.9in]{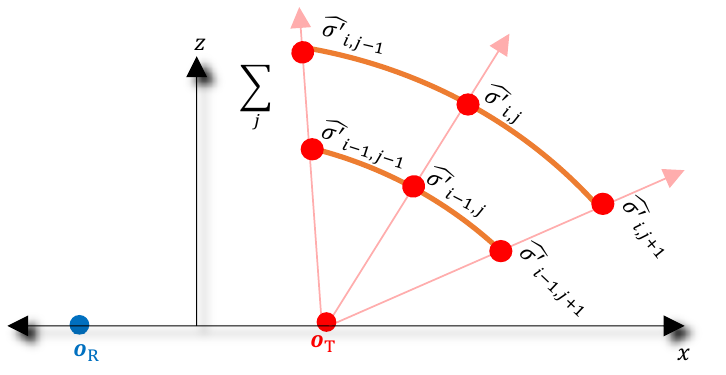}
  \caption{2D diagram ($xz$ slice) of our scheme for integrating ellipsoid surfaces to synthesize measurements. Eq.~\eqref{eq:data-consistency} weights scene scatterers by their respective transmission and Lambertian terms and integrates them along the surface of ellipsoids.}
  \label{fig:ellipse-integrate}
\end{figure}

\subsection{Transmission and Normal Calculations}
\label{sec:occ-normal}
We handle occlusion by computing transmission probabilities between the transmitter/receiver and scene points. Our computed transmission probability is similar to NeRF's~\cite{mildenhall2021nerf}, although we use it to weight complex-valued scatter coefficients rather than scene density.


As shown in Fig.~\ref{fig:ellipse-integrate}, the network $N_{\text{BP}}$ predicts complex-valued scatterers $(\netout_{ij})$ at the sampled ray-ellipsoid intersections. The transmission probability between the transmitter and scene point is computed using the cumulative product in depth~\cite{mildenhall2021nerf}
\begin{equation}
    T(\mathbf{o}_T, \mathbf{x}_{T_i}) = \prod_{k < i}\exp{\left(-|\netout_k|\cdot|l_{k+1} - l_k|\cdot\zeta\right)},
    \label{eq:tx-prob}
\end{equation}
where $k$ indexes depth, and we omit the direction index $j$ since Eq.~\eqref{eq:tx-prob} is computed for all rays. We use the scalar $\zeta$ to scale the transmission falloff rate --- we find it useful to increase $\zeta$ for sparser pulse deconvolved waveforms (corresponding to sparser scenes). Note that we use the scatterer magnitude to compute the transmission probability since each term in the cumulative product should be non-negative.

Computing a return ray from each sampled transmission point approximately squares the number of required scene samples. Thus, we compute a return ray only from the expected depth of the transmission ray \cite{zhang2021nerfactor}.  The return ray is illustrated in blue in Fig.~\ref{fig:sampling-scheme}. We compute the expected ray sample in-depth as,
\begin{equation}
    \mathbb{E}(\mathbf{x}_T) = \sum_i \mathbf{x}_{T_i} \frac{|\netout(\mathbf{x}_i) T(\mathbf{o}_T, \mathbf{x}_{T_i})|}{\sum_i|\netout(\mathbf{x}_i)T(\mathbf{o}_T, \mathbf{x}_{T_i})|}.
    \label{eq:expected-range}
\end{equation}
and define the return ray as
\begin{equation}
    \mathbf{x}_{R_{ij}} = \mathbb{E}(\mathbf{x}_{T_i}) + l_i \mathbf{d}_{R_j},
\end{equation}
where $\mathbf{d}_{R_j}$ = $\frac{\mathbf{o}_R - \mathbb{E}(\mathbf{x}_{T_{j}})}{||\mathbf{o}_R - \mathbb{E}(\mathbf{x}_{T_{j}})||}$ and the depths $l_i$ are sampled at the ellipsoid intersections found using the negative root of Eq.~\eqref{eq:quadratic}. Since the expected depth is typically less than the max depth (see Fig.~\ref{fig:sampling-scheme}), we simply set return ray samples with depths greater than the expected sample (i.e. when $l < 0$) to 0 so that they are ignored by downstream calculations. We compute a transmission probability for the return ray using a cumulative product in depth (Eq.~\ref{eq:tx-prob}). The return ray is used only to compute the transmission probability --- its sampled points are not part of the ellipsoid surface integrations of Eq.~\eqref{eq:data-consistency}. Note that for simulated and real AirSAS experiments, the transmitter and receiver are collocated so we omit calculating the transmittance of the returning ray.

We compute surface normals using a method from Poole et al.~\cite{poole2022dreamfusion},
\begin{equation}
     \mathbf{n}(\mathbf{x}) = \frac{\nabla \mathbf{x}|\netout|}{\left|\left|\nabla \mathbf{x}|\netout|\right|\right|},
     \label{eq:normals}
\end{equation}
noting that we use the magnitude of the scatterers for normal computation.
Scatterers are weighted with a Lambertian scattering model model~\cite{lambert1760photometria, ramamoorthi2001signal}
\begin{equation}
    L(\netout) = \netout \cdot \max\left(0, \mathbf{n}(\mathbf{x}) \cdot \frac{\mathbf{x} - \mathbf{o}_T}{\left|\left|\mathbf{x} - \mathbf{o}_T\right|\right|}\right).
\end{equation}
We show in our experiments that the Lambertian scattering model is important for reconstructing accurate object surfaces. 

\subsection{Loss and Regularization} 
\label{sec:regularization}

Fig.~\ref{fig:ellipse-integrate} illustrates integrating sampled ellipsoid surfaces after weighting scene scatterers with transmission and Lambertian terms. As expressed in Eq.~\eqref{eq:data-consistency}, these operations synthesize a complex-valued waveform. We compute a loss between the synthesized waveform and the analytic version of the pulse deconvolved waveforms,
\begin{align}
\begin{gathered}
    \mathcal{L}_{BP} = \left|\left|\widehat{\mathbf{s'}_{\text{PD}}} - \widehat{\mathbf{s}_{\text{PD}}}\right|\right|_2 + \\ \lambda_{1_{\text{BP}}}\mathcal{L}_{\text{Sparse}}^{\text{BP}} + \lambda_{2_{\text{BP}}}\mathcal{L}_{\text{TV}_{\text{Space}}}^{\text{BP}} + \lambda_{3_{\text{BP}}}\mathcal{L}_{\text{TV}_{\text{Phase}}}^{\text{BP}} \label{eq:a-by-s-bp}
\end{gathered}
\end{align}
where $\lambda_{1_{\text{BP}}}$, $\lambda_{2_{\text{BP}}}$, and $\lambda_{3_{\text{BP}}}$ are scalar weights for commonly used sparsity and total variation priors defined as,
\begin{align}
    \mathcal{L}_{\text{Sparse}}^{\text{BP}} = \sum_n \left|\left||\netout|\right|\right|_1 \\
    \mathcal{L}_{\text{TV}_{\text{Space}}}^{\text{BP}} = \sum_n||\nabla_{d_{\text{reg}}} \netout||_1, \\
    \mathcal{L}_{\text{TV}_{\text{Phase}}}^{\text{BP}} = \sum_n||\nabla_{d_{\text{reg}}} \angle \netout||_1.
\end{align}
The total loss is minimized with respect to the network weights. The total variation losses are performed on the complex scene scatterers and their phase --- the $\nabla_{d_{\text{reg}}}$ is computed using the distance hyperparameter $d_{\text{reg}}$ that determines the distance between the compared samples. Regularization terms are computed using all ray samples. We find that these priors benefit some reconstructions while harming others and should be tuned depending on the scene and measurement noise. For example, in the supplemental material we show that these priors are particularly useful when reconstructing from limited measurements. We highlight that the ability to add and tune priors is an advantage of our method over backprojection, which does not have the ability to incorporate priors~\cite{ahn2019convolutional}. 

Eq.~\eqref{eq:a-by-s-bp} should be minimized over the given $n$ sensor measurements. In practice, we use gradient accumulation to average the gradients over a fixed number of sensors before performing a backpropagation update to the weights. We find that this stabilizes the optimization while avoiding the memory overhead of batching multiple sensors. In all experiments, we accumulate gradients over $5$ sensors.

Finally, we note that the loss function as shown in Eq.~\eqref{eq:a-by-s-bp} can be considered coherent since it computes the loss between the complex estimated and target deconvolved measurements. In the supplemental material, we show that incoherent reconstruction yields inferior results, validating our design choice of having the network predict complex-valued scatterers and performing the loss between complex-valued measurements.

\begin{figure}
  \centering
  \includegraphics[width=3.35in]{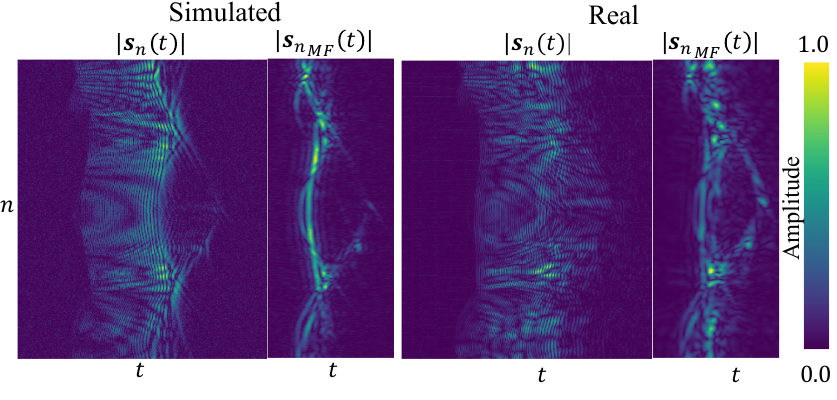}
  \caption{Comparing simulated measurements and real measurements of the bunny measured using an LFM pulse ($f_c=20$~kHz and $\Delta f = 20$~kHz). Our simulated measurements are qualitatively similar to real measurements.}
  \label{fig:sim-real-comparison}
\end{figure}

\begin{figure}
  \centering
  \includegraphics[width=3.3in]{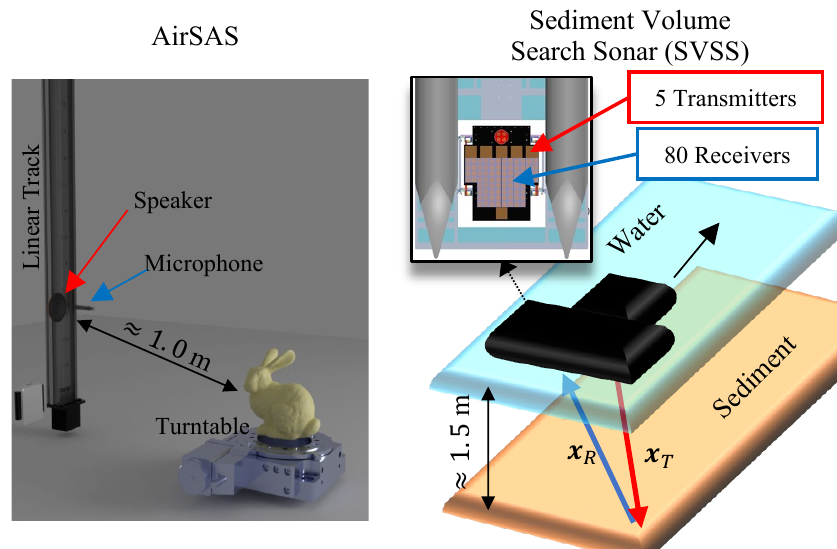}
  \caption{Measurement geometry for the AirSAS (left) and SVSS (right). AirSAS collects circular SAS measurements inside an anechoic chamber using a speaker and microphone pointed at a turntable. The speaker and microphone are mounted on a vertical linear track $1$ meter from the turntable. Objects on the turntable fit within 0.2 cubic meters. SVSS is mounted on a pontoon boat that travels in a linear fashion and is used to image underwater targets. The SVSS array is approximately $1.5$ meters above the lakebed and reconstructs scenes with dimension approximatley $1$ meter across (perpendicular to platform path) and between $1-10$ meters long (parallel to platform path).}
  \label{fig:airsas-svss-schematic}
\end{figure}

\section{Simulator and Hardware Implementation}
\label{sec:implementation}

In this section, we discuss the implementation of a simulator and hardware for collecting SAS measurements. We refer the reader to open-source code and data.\footnote{Code, AirSAS, and simulation data are available through the website \href{https://awreed.github.io/Neural-Volume-Reconstruction-for-Coherent-SAS/}{(click here)}. For SVSS data, readers must contact Daniel C. Brown (dcb19@psu.edu) for permission to use per funding agency guidelines. Refer to this link \href{https://serdp-estcp.org/projects/details/97e4a43c-bb65-432d-94ba-6aaff361571a}{(click here)} for more SVSS details.}

\subsection{SAS Simulator using Optical Time-of-Flight Renderer}

We simulate SAS measurements by leveraging an implementation of an optical time-of-flight (ToF) renderer based on Kim et al.~\cite{kim2021fast}. This renderer was chosen in part due to its CUDA implementation that uses asynchronous operations to efficiently accumulate per-ray radiances by their travel time. While this simulator does not capture acoustic effects (including diffraction), it does enable efficient prototyping and we note that optical renderers have been leveraged for SAS simulation in the literature~\cite{reed2019coupling}. 

Specifically, we configure the simulator to emulate an in-air, circular SAS setup called AirSAS~\cite{Blanford2019}. AirSAS consists of a speaker and microphone directed at a circular turntable that holds a target. We simulate the speaker (transmitter) as a point light source and the microphone (receiver) with an irradiance meter to measure reflected rays. We configure the renderer to measure the ToF transients from each sensor location. We convolve these transients with our transmitted pulse to obtain simulated SAS measurements. We provide plots of rendered transient signals in the supplemental document.

We use simulated measurements for several experiments to quantitatively evaluate the effect of bandwidth, noise, and object shape. In Fig.~\ref{fig:sim-real-comparison}, we show a side-by-side comparison of a subset of simulated and measured waveforms from the bunny. While our simulator ignores non-linear wave effects like sound diffraction and environmental effects like changing sound speed, we observe that the simulated measurements are similar to AirSAS measurements.

\subsection{AirSAS}
\label{subsec:airsas}
AirSAS is an in-air, circular SAS contained within an anechoic chamber~\cite{Blanford2019}. AirSAS being an in-air system enables experimental control that is impossible or challenging to achieve in water. Importantly, the relevant sound physics between air and water are directly analogous for the purposes of this work. AirSAS data has been used extensively in prior literature for proof-of-concept demonstrations~\cite{park2020alternative,cowen2021airsas, goehle2022approximate, blanford2022leveraging, SINR}. 

We illustrate the concept of AirSAS on the left side of Fig.~\ref{fig:airsas-svss-schematic}. The AirSAS transducer array is comprised of a loudspeaker tweeter (Peerless OX20SC02-04) and a microphone (GRAS 46AM). The tweeter transmits a linear frequency modulated chirp for a duration of 1 ms waveform at center frequency $f_c=20$ kHz and bandwidth $\Delta f=20$ kHz or $\Delta f=5$ kHz depending on the experiment. The transmitted LFM is multiplied with a Tukey window with ratio $0.1$ to suppress side-lobes. 
 
AirSAS measurements of 3D printed objects (shown in Fig.~\ref{fig:teaser}) were provided by the authors of~\cite{Blanford2019}. AirSAS scenes are collected on a $0.2 \times 0.2$ meter turntable that is rotated in 360, $1$ degree increments. The speaker and microphone are placed approximately $1$ meter from the table and vertically translated by a linear-track by $5$ mm at every 360 measurements. The spacing between measurements satisfies SAS sampling criteria of $D \leq \lambda_{\text{min}}/2$ where $\lambda_{\text{min}}$ is the minimum wavelength in the transmit waveform and $D$ is the distance between measurement elements~\cite{Callow2003}. We can easily sub-sample the full set of given measurements to create helical or sparse-angle collection geometries (e.g., those used in Fig.~\ref{fig:undersampled-results}).

\subsection{Sediment Volume Search Sonar} 
We also evaluate our method on in-water sonar measurements collected from the Sediment Volume Search Sonar (SVSS)~\cite{svss-tech-report}. SVSS was designed for sub-surface imaging and thus uses relatively long wavelengths to penetrate a lake bed. Specifically, the array transmitters emit an LFM with spectra ($f_c = 27.5$ kHz and $\Delta f = 15$~kHz) for a duration of $255$~$\mu$s and Taylor windowed~\cite{svss-tech-report}. The SVSS is deployed on a pontoon boat (shown in Fig.~\ref{fig:teaser}) that is equipped with a suite of precise navigation sensors to accurately tow the sonar array in the water. We were provided with field data from the Foster Joseph Sayers Reservoir in Pennsylvania where various targets and objects of interest were placed on the lakebed and then measured~\cite{svss-tech-report}. 

Fig.~\ref{fig:airsas-svss-schematic} shows the SVSS array and the measurement geometry. The SVSS transducer array consists of 5 transmitters that ping the scene in cyclical succession and 80 actively recording receive elements. For our reconstructions, we typically discard measurements where the scene target is outside the beamwidth of the firing transmitter. Unlike AirSAS, we do not assume a collocated transmitter and receiver for SVSS --- the transmit and receive elements are relatively far apart relative to the imaging range (i.e., bistatic). We account for the bistatic transmit and receive elements by computing a return ray for each transmit ray at the expected transmission depth as described in Section~\ref{sec:occ-normal}.

\subsection{Methods for Comparison}
We compare our method to two 3D SAS imaging algorithms: time-domain backprojection and a polar formatting algorithm (PFA). We use traditionally matched-filtered waveforms as input to the SAS imaging algorithms, except for the ablation experiment shown in Figure~\ref{fig:quad-ablation}. Time-domain backprojection focuses the matched-filtered waveforms onto the scene by explicitly computing the delay between the sensor and scene. Backprojection applies to near arbitrary array and measurement trajectories~\cite{callow2009motion} and is standard for high-resolution SAS imaging~\cite{hansen2011challenges, gerg2020gpu}, making it the stable baseline~\cite{callow2006effect} for our AirSAS and SVSS experiments. For breadth of comparison, we also implement and compare against a polar formatting algorithm (PFA)~\cite{gough1997unified}, a wavenumber method designed for circular SAS. This algorithm applies to our circular AirSAS and simulation geometries, but not the non-linear and bistatic measurement geometry of SVSS~\cite{Callow2003}.

Note that there are several existing analysis-by-synthesis reconstruction methods for 2D SAS, but these are not easily adapted to 3D SAS with non-linear and bistatic measurement geometries. Reed et al. present a deconvolution method for 2D SAS that assumes a spatially invariant PSF, does not consider 3D effects like occlusion and surface normals, and is applies only to 2D circular SAS~\cite{SINR}. Putney and Anderson adapt the WIPE deconvolution method for 2D SAS~\cite{Putney2005}. While WIPE was originally designed for 1D deconvolution~\cite{lannes1997clean}, Putney and Anderson extend it to 2D SAS by inverting the range migration algorithm (RMA) and evaluate their method on two SAS images~\cite{Putney2005}. The paper does not provide reproducible details on how to invert the RMA, how to apply it to SAS, or provide available code/software. Future work is needed to implement and adapt WIPE to consider 3D effects and complexities such as occlusion, surface scattering, bistatic arrays, and arbitrary collection geometries.

In addition to backprojection and PFA, we implement a `Gradient Descent' baseline, which is neural backprojection method without the INR. Instead of an INR, we backpropagate error gradients directly to a fixed set of scene voxels. The value of a scene point is computed by trilinear interpolation of the relevant voxels. This comparison is similar to Yu et al., who implement NeRF without a neural network~\cite{yu_and_fridovichkeil2021plenoxels}, but we note that they optimize a spherical harmonic basis rather than scene values directly. In our case, the gradient descent comparison is simply the removal of the INR from neural backprojection, allowing us to better observe the INR's specific impacts on reconstruction quality.





\subsection{Optimization, Visualizations, and Metrics}
We reconstruct all real results using an A100 GPU and simulated results using a 3090 Ti GPU. We use PyTorch's python library for all experiments. For pulse deconvolution, we choose an INR hash encoding and model architecture derived from `Instant NeRF' by~\cite{mueller2022instant} for its convergence speed. For neural backprojection, we use the same hash encoding technique from~\cite{mueller2022instant} coupled with four fully-connected layers. Using the A100 system, it takes approximately 40 ms to deconvolve a single 1000 sample measurement. For reconstruction, using 5000 rays and 200 depth samples, it takes our proposed method approximately 10 ms per iteration, and approximately 10,000 - 20,000 iterations to converge. The gradient descent method runs marginally slower at approximately 16 ms per iteration. The number of iterations until convergence was approximately equivalent for all scenes reconstructed using between 2000 and 50000 measurements. Finally, we note that backprojection is faster than our iterative methods as it takes approximately 0.1 ms per measurement (analogous to one iteration since we process one measurement per iteration). Overall it takes approximately 1-2 hours to reconstruct AirSAS and SVSS scenes with our method or gradient descent while backprojection of these scenes takes less than 5 minutes. 



We visualize AirSAS and SVSS reconstructions using MATLAB's volumetric renderering function, \texttt{volshow()}. Please refer to the supplemental material for AirSAS renderings at different thresholds. For the SVSS data, we also use maximum intensity projections (MIPs) to better visualize the data collapsed into two dimensions~\cite{Wallis} and more easily measure the dimensions of reconstructed targets in the supplemental material. For simulated visualizations, we use marching cubes to export a mesh and render depth and illumination colored images. 
We show all methods on the same threshold to provide the reader with a fair comparison. 

\begin{figure}[b]
  \centering
  \includegraphics[width=3.5in]{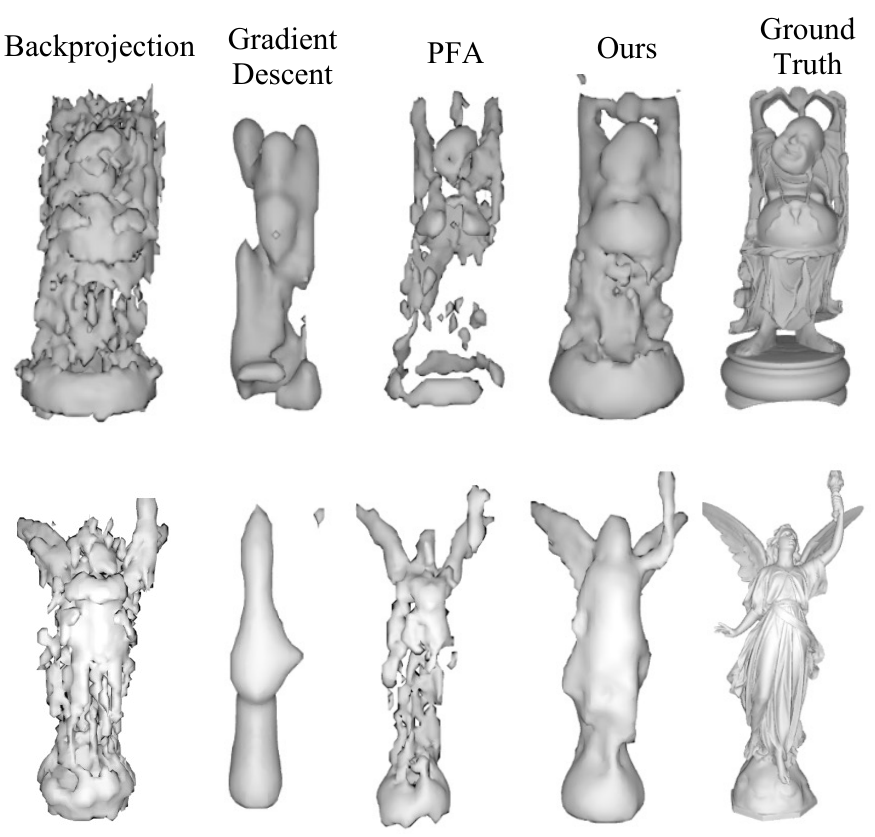}
  \caption{Reconstructions from simulated measurements using backprojection, gradient descent, the polar formatting algorithm (PFA), and our method. Compared to other methods, our reconstructions more accurately match the ground truth geometry.}
  \label{fig:more_meshes_paper}
\end{figure}

Meanwhile, to quantitatively evaluate each method, we use two 3D metrics (Chamfer~\cite{borgefors1986distance} distance, intersection-over-union (IOU)) and two image metrics (PSNR, LPIPS~\cite{zhang2018unreasonable}) for selected viewpoints of the 3D volume. The 3D metrics capture the entire point cloud reconstruction performance, while the 2D image metrics help capture the perceptual quality of the shape from the rendered viewpoints. We compute points for the 3D metrics by exporting each method's predicted point cloud to a mesh. As the point cloud threshold (i.e., points < threshold = 0) influences the predicted mesh, we sweep over possible thresholds and choose the threshold that maximizes the performance of each method. The Chamfer distance is calculated based on a point cloud sampled from the reconstructed mesh surface, while IOU is calculated using a voxel representation of the mesh. We compute image metrics on rendered depth-images of the predicted and ground truth mesh at 10 azimuth angles. We use the depth images as these are independent on illumination and rely solely on the the object's reconstructed geometry.



\begin{table}[]
\caption{Simulation results showing the average quantitative metrics for Backprojection (BP), Gradient descent (GD), the Polar formatting algorithm (PFA), and our reconstructions of 8 different meshes.}
\begin{tabular}{llllll}
\hline
Metric           & Chamfer $\downarrow$           & IOU $\uparrow$            & LPIPS  $\downarrow$         & PSNR $\uparrow$   & MSE $\downarrow$           \\

\hline
BP   & 1.36E-04          & 0.2928          & 0.1215          & 15.783        & 5.55E-03 \\
GD & 2.21E-04          & 0.4309          & 0.1236          & 15.117        & 6.13E-03 \\
PFA             & 2.13E-04          & 0.3586         & 0.1238          & 15.048          & 6.64E-03 \\
Ours             & \textbf{1.12E-04} & \textbf{0.5194} & \textbf{0.0988} & \textbf{17.918} & \textbf{3.99E-03} \\
\hline
\end{tabular}
\label{tab:mesh_averaged_result}
\end{table}

\begin{figure}
  \centering
  \includegraphics[width=3.4in]{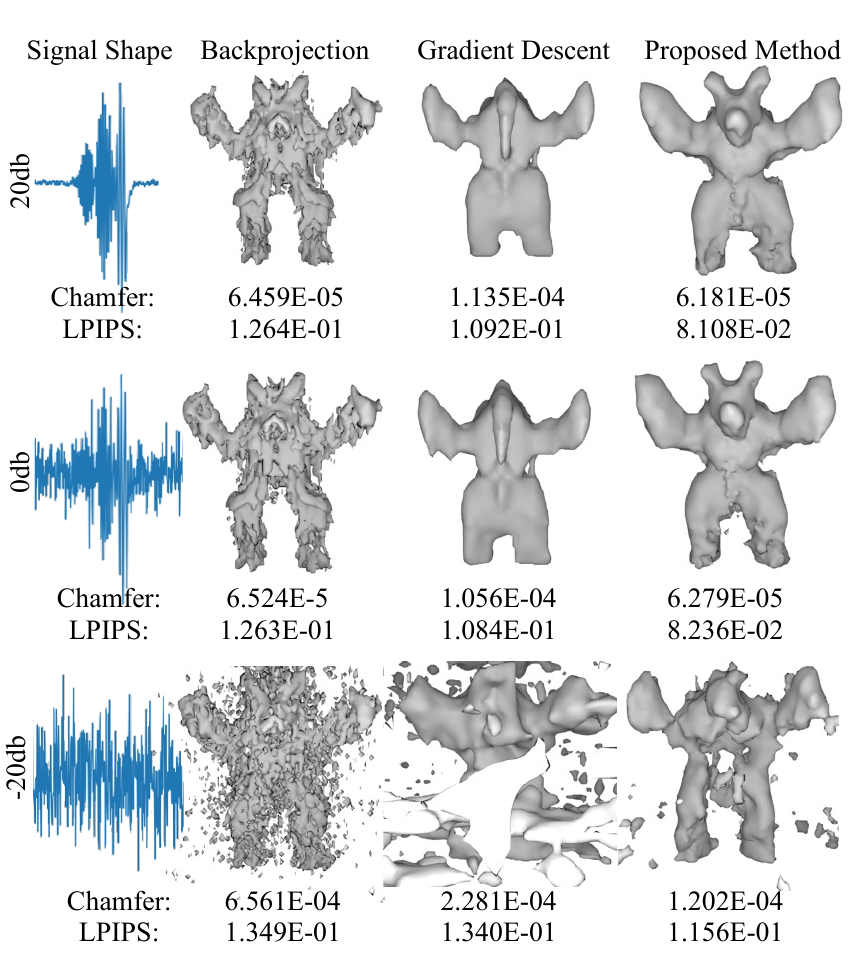}
  \caption{Simulation results using a $\Delta f = 20$ kHz LFM showing the reconstructed meshes of an armadillo object at three noise levels. Our method performs decently well even at -20 dB signal-to-noise-ratio. }
  \label{fig:armadilo_noise}
\end{figure}

\begin{figure}[h!]
  \centering
  \includegraphics[width=3.4in]{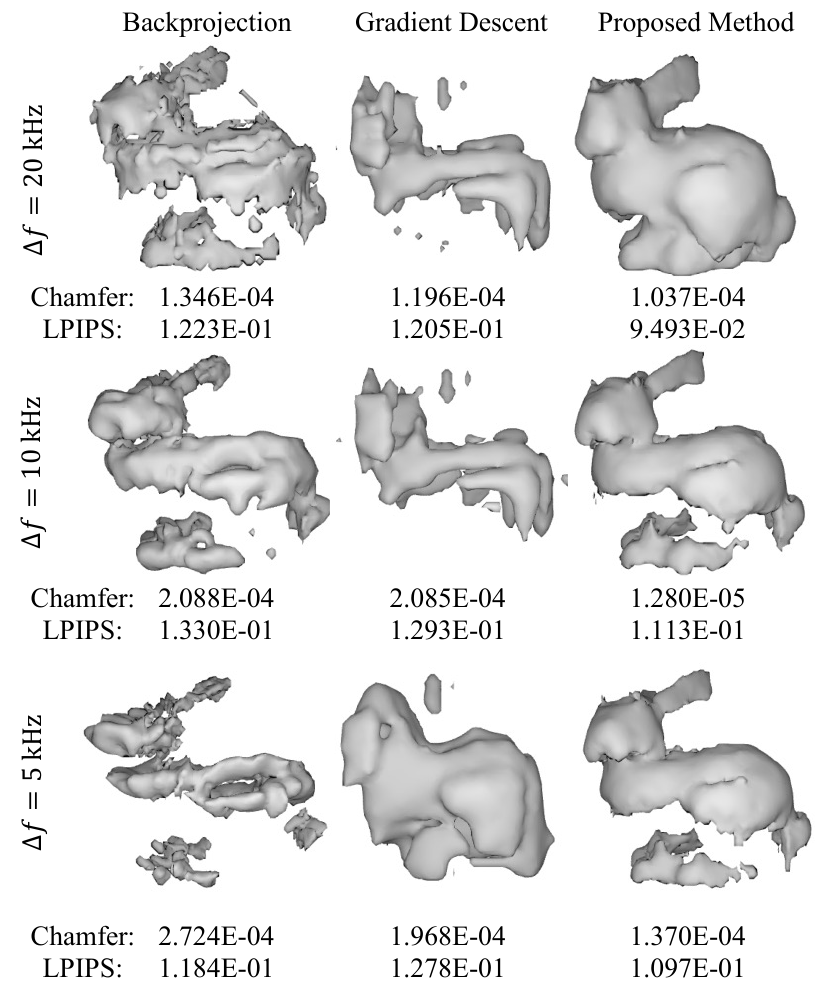}
  \caption{Simulation results using LFMs with different bandwidths $\Delta f$ to measure a bunny. As expected, the performance of all methods degrades at lower bandwidths, but our proposed method better preserves the object geometry.}
  \label{fig:bunny-bandwidth}
\end{figure}

\section{Experimental Results}
\label{sec:experimentalresults}

We validate our method on simulated data and two real-data sources. Section~\ref{sec:simulation_results} presents our simulation results, where we test our method against baselines while varying experiment noise and bandwidths. Section~\ref{sec:airsasresults} presents our first real-data source, AirSAS, where we test our method under varying measurement trajectories, bandwidths, and ablations. Finally, Section~\ref{sec:svssresults} presents our other real-data source, measurements captured of the Foster Joseph Sayers Reservoir, Pennsylvania using SVSS. Our SVSS results validate our method's applicability to underwater environments and bistatic transducer arrays.


\begin{figure*}
  \centering
  \includegraphics[width=5.0in]{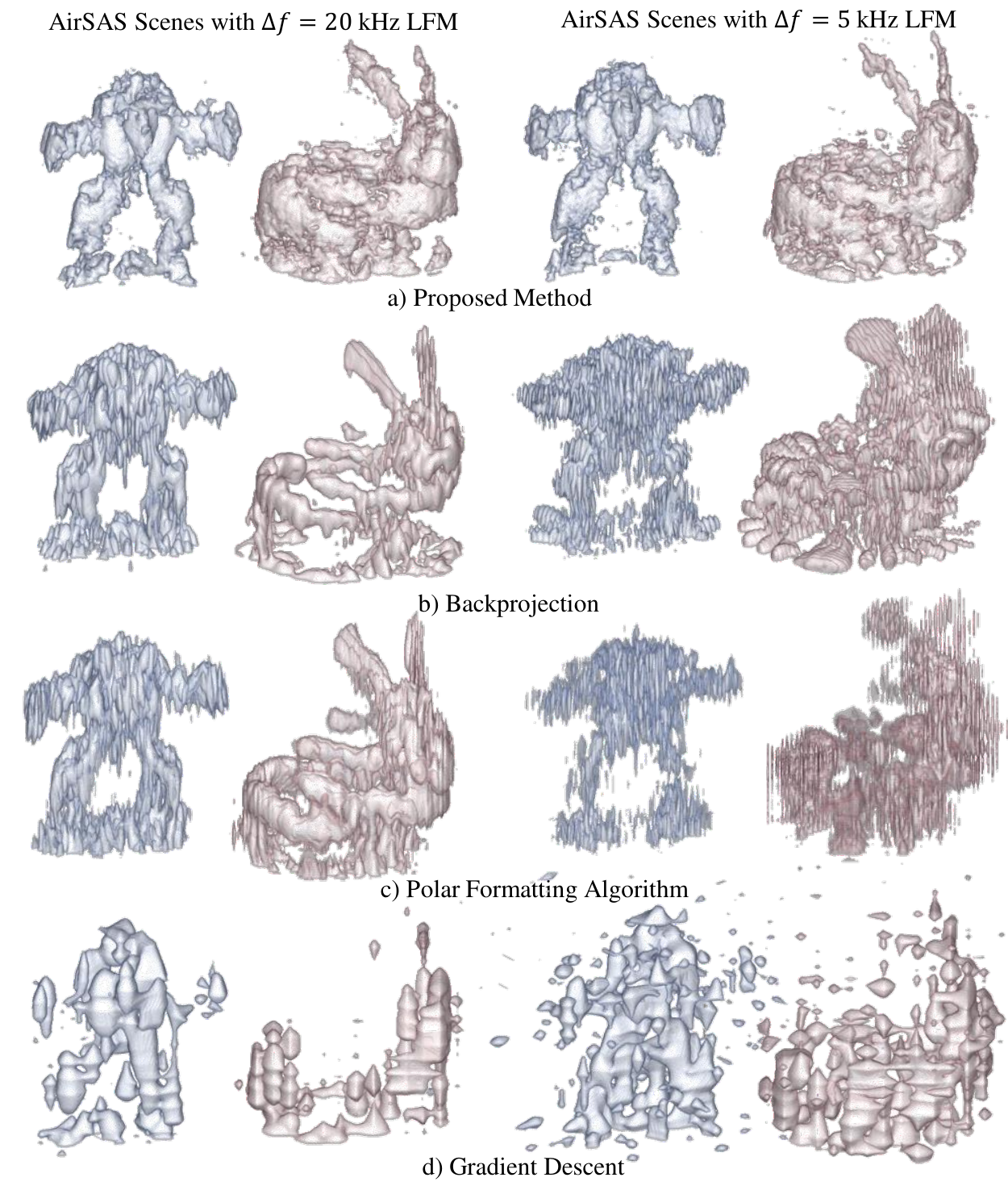}
  \caption{Reconstructions of AirSAS data captured with relatively high ($\Delta f = 20 $ kHz) and low ($\Delta f = 5 $ kHz) bandwidth LFMs. Our method demonstrates more consistent performance across waveform bandwidth compared to backprojection, the polar formatting algorithm, and gradient descent.}
  \label{fig:airsas-results}
\end{figure*}

\subsection{Simulation Results}
\label{sec:simulation_results}
\paragraph{Effectiveness of proposed method}
We compare our proposed method against backprojection, the polar formatting algorithm, and gradient descent on simulated scenes measured with an LFM ($f_c = 20$ kHz and $\Delta f = 20$~kHz) at an SNR of $20$~dB. Table~\ref{tab:mesh_averaged_result} presents quantitative metrics averaged over reconstructions from 8 different objects. Considering the PSNR metric, our approach offers $2$ dB improvement over other methods. 

Figure~\ref{fig:more_meshes_paper} shows reconstructions of two objects used to compute the average in Table~\ref{tab:mesh_averaged_result}. In general, our method more accurately matches the ground truth geometry, especially occlusions in the mesh. We note that our method does lose some high-frequency details. We show reconstructions for all 8 meshes in the supplemental material. 





\paragraph{Effects of noise: } In Fig.~\ref{fig:armadilo_noise}, we show armadillo reconstructions from each method at three SNR levels and measured using a $\Delta f = 20$~kHz LFM. The bottom row of the figure shows an example waveform to highlight the challenge of reconstructing the scene under poor SNR conditions. As expected, the performance of each method tends to degrade as SNR decreases. Our method tends to outperform both competing methods at each noise level\footnote{We do not include the PFA algorithm in these simulation experiments since it underperforms backprojection.}. We observe that the gradient descent method fails to recover higher frequency details on the object. There also seems to be a saturation level where 20 dB is not that much improved over 0 dB for all methods.  


\paragraph{Effects of bandwidth:}
Fig.~\ref{fig:bunny-bandwidth} shows our next experiment, where a bunny object is simulated using three LFM bandwidths, $\Delta f = 5kHz$, $\Delta f = 10kHz$, and $\Delta f = 20kHz$. As expected, the performance of all methods degrades at lower bandwidths, but we observe that our method tends to preserve the shape of spatial features more accurately compared to backprojection and gradient descent. This observation is reflected in the quantitative results, where our reconstructions generally report superior metrics.

\begin{figure*}[t]
  \centering
  \includegraphics[width=7.0in]{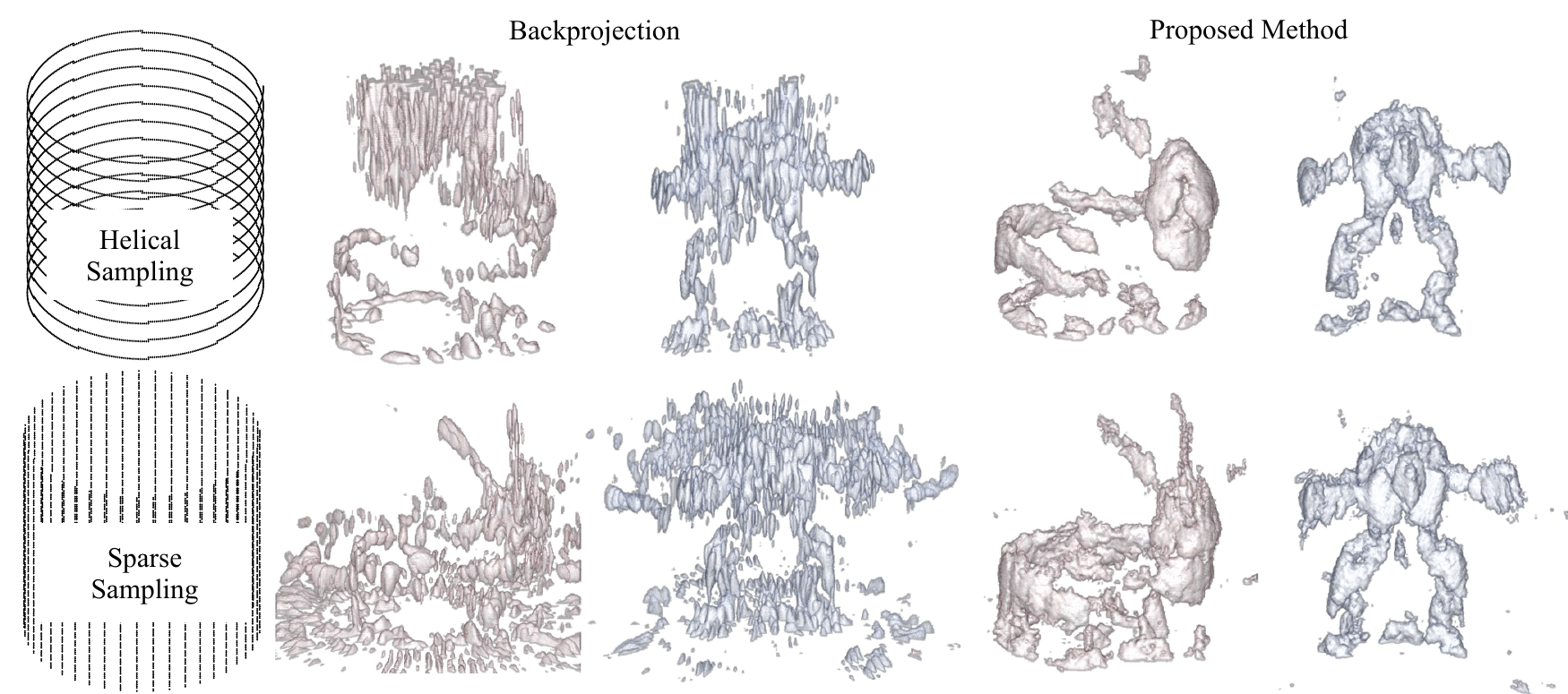}
  \caption{Helical and sparse view reconstructions of AirSAS measurements from $\Delta f = 20$ kHz AirSAS bunny and armadillo. Our method can recover scene geometry while attenuating the undesirable image artifacts that plague backprojection in undersampled regimes.}
  \label{fig:undersampled-results}
\end{figure*}

\subsection{AirSAS Reconstructions}
\label{sec:airsasresults}

\paragraph{Main results:} We reconstruct a 3D printed bunny and armadillo that were measured with an LFM of center frequency $f_c=20$kHz at bandwidths $\Delta f = 20$~kHz  and $\Delta f = 5$~kHz. Fig.~\ref{fig:airsas-results} shows the results of our method, backprojection, the polar formatting algorithm and gradient descent. First, we note that reconstruction quality decreases across all methods as the bandwidth decreases. However, the quality of our method is notably more consistent across the two bandwidth values, which conforms to our simulated results presented earlier. In the lower bandwidth case, our method produces notably more detailed and artifact-free reconstructions compared to the other methods.  This is partially due to the fact that our pulse deconvolution performs similarly across both bandwidths, which is not the case for backprojection since it uses matched filtered waveforms. Additionally, neural backprojection helps overcome structural errors present in backprojection and PFA, such as the bunny's concave head at the $20$~kHz bandwidth. We observe that the gradient descent method's reconstruction is noisier and less accurate than ours, highlighting the importance of using an INR in neural backprojection.

\paragraph{Undersampling: } In real-world SAS applications, it is difficult to obtain multiple looks at an object from a dense collection of viewpoints. Thus we compare the performance of our method to backprojection in helical and sparse sampling schemes, as shown in Fig.~\ref{fig:undersampled-results}. In both the helical and sparse sampling cases, we are using only approximately $10\%$ of the measurements required to be fully sampled in the traditional sense. Helical sampling is missing many vertical samples, and therefore induces vertical streaking artifacts in the backprojection results, as discussed in other works~\cite{marston2016volumetric}. While our method's reconstruction is not perfect in this case, we highlight the fact that it contains fewer vertical streaking artifacts when compared to backprojection. 

Sparse view sampling is common in computed-tomography literature, and is known to induce radial streaking artifacts in imagery due to the missing angles~\cite{bian2010evaluation}. As shown in the bottom row of Fig.~\ref{fig:undersampled-results}, our method does remarkably well in this case, reconstructing a scene that is comparable to one obtained with fully sampled measurements. The notably superior performance of our method can potentially be attributed to our sparsity and smoothness priors, and also aligns with previous works that demonstrate the utility of INRs in limited data reconstruction problems~\cite{ruckert2022neat, reed4Dct, sun2022coil}.

\begin{figure}
  \centering
  \includegraphics[width=2.75in]{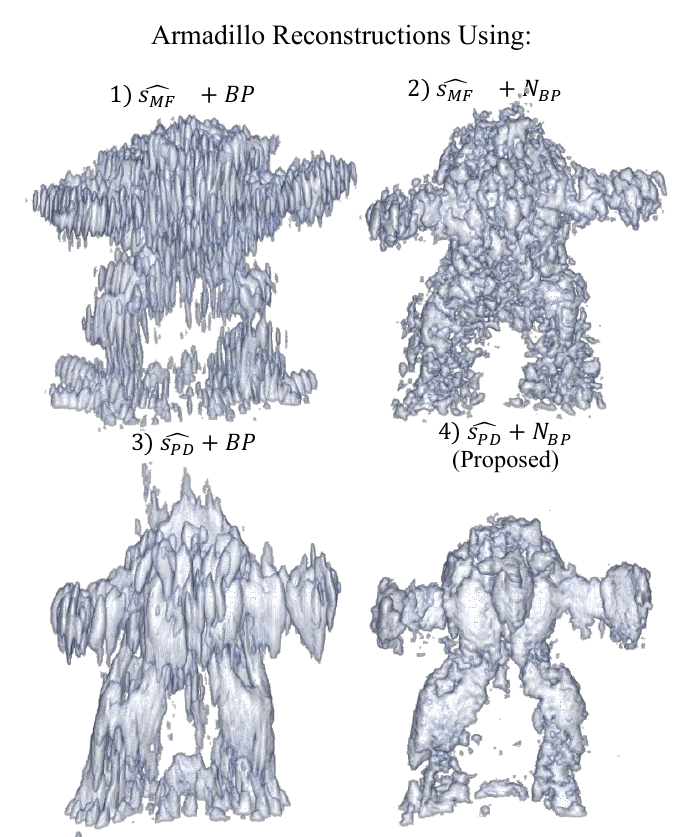}
  \caption{Ablating pulse deconvolution and neural backprojection using AirSAS $\Delta f = 5$~kHz armadillo measurements. The top left (1) shows the traditional reconstruction pipeline that uses matched filtered waveforms as input to backprojection. The top right (2) shows using the matched filtered waveforms as input to neural backprojection. The bottom left (3) shows using the pulse deconvolved measurements as input to backprojection. The bottom right (4) is our proposed method and uses pulse deconvolution as input to neural backprojection. We observe that pulse deconvolution helps attenuate high-frequency surface noise, while neural backprojection aids in reconstructing accurate geometric features. Both pulse deconvolution and neural backprojection are important for maximizing reconstruction performance.}
  \label{fig:quad-ablation}
\end{figure}

\paragraph{Ablation studies:}
Our ablations investigate various design choices in our pipeline. First, we perform an experiment to highlight the synergistic importance of both pulse deconvolution and neural backprojection in our method. In Fig.~\ref{fig:quad-ablation}, we show an armadillo reconstruction using (1) matched-filtering and time-domain backprojection, (2) using matched-filtering and neural backprojection, (3) pulse deconvolution and time-domain backprojection, and (4) pulse deconvolution and neural backprojection (proposed approach). Clearly, both steps of our method are important for maximizing reconstruction quality. We observe that using match filtered waveforms with neural backprojection gives perhaps a slightly more accurate geometry, but is extremely noisy due to the limited range compression abilities of match-filtering. On the other hand, using our pulse deconvolved waveforms with  traditional backprojection yields a smoother reconstruction than the traditional pipeline, but contains streaking artifacts common in backprojection algorithms.

\begin{figure}
  \centering
  \includegraphics[width=3.35in]{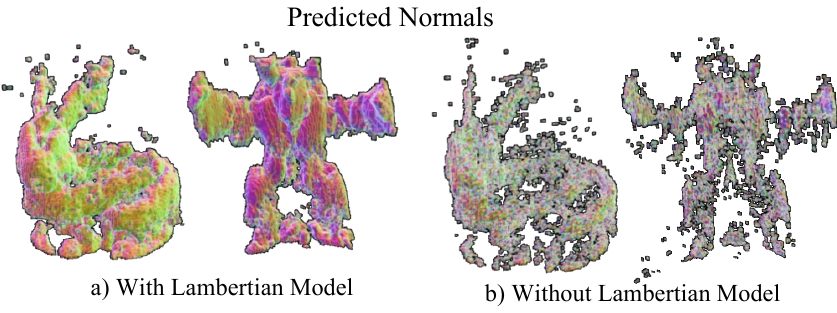}
  \caption{AirSAS reconstructions of scenes measured with a $\Delta f = 20$ kHz LFM weighted by the 3D predicted normals encoded using the RGB channels. Left (a) shows reconstructions using the Lambertian model and right (b) shows reconstructions without using the Lambertian model. Using the Lambertian model yields better reconstructions and normal predictions since it enables the model to account for incidence angle.}
  \label{fig:ablation-normals}
\end{figure}

Fig.~\ref{fig:ablation-normals} demonstrates the importance of our Lambertian scattering model by visualizing the normals computed by Eq. ~\eqref{eq:normals}. The reconstruction without the Lambertian model is noisier and sparser as the network struggles to reconstruct a scene consistent with given measurements since the 3D printed object is roughly diffuse in acoustic scattering. The normals also appear almost random without a Lambertian model, as the network was not constrained to output consistent surfaces. 

\begin{figure}
  \centering
  \includegraphics[width=3.5in]{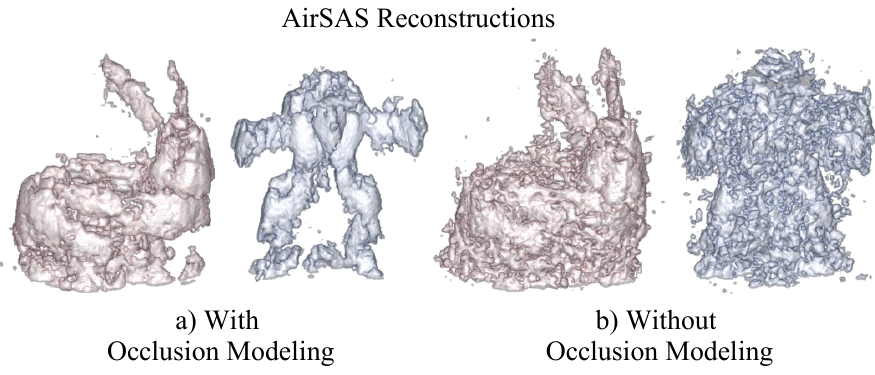}
  \caption{AirSAS reconstructions of scenes measured with a $\Delta f = 20$ kHz LFM (a) with occlusion modeling and (b) without occlusion modeling. We turn off occlusion by setting $\zeta=0$ in the transmission probability of Eq.~\ref{eq:tx-prob}. Occlusion modeling is important for our method to obtain accurate reconstructions.}
  \label{fig:ablation-occlusion}
\end{figure}

Fig.~\ref{fig:ablation-occlusion} ablates the ability of our forward model to handle occlusion by setting $\zeta=0$ in the transmission probability calculation of Eq.~\ref{eq:tx-prob}. Without occlusion, neural backprojection fails to capture sharp outlines of the object's geometry. As occlusion was a factor during the capture of the real-data, these artifacts are expected --- the network predicts erroneous features since the forward model is unable to account for occlusion.

In the supplemental material, we also ablate for coherent versus incoherent SAS reconstruction as well as the effect of scene priors (e.g., sparsity) on reconstructions.

\subsection{SVSS Reconstructions}
\label{sec:svssresults}
Migrating from an in-lab to an in-water SAS deployed in the field brings a new set of challenges. First, the SVSS uses a more complicated sonar array that consists of five transmitters and eighty receivers, each with overlapping beamwidth that should be considered for accurate reconstructions. 

Additionally, the energy backscattered from the lakebed is relatively strong compared to the targets. As such, we observe that a naive application of our deconvolution method using sparsity regularization tends to deconvolve returns from the lakebed and set the significantly smaller energy from the target close to zero. Using these measurements with neural backprojection yields subpar results since the objective function is mainly concerned with reconstructing the background. We address this issue by dynamic-range compressing our deconvolved measurements before passing them to the network --- while this step amplifies noise, we find that that it makes the energy from the target strong enough for quality reconstructions. In particular, we dynamic-range compress measurements using $\mathbf{s}_{\text{PD}}^{\text{drc}} = \text{sign}(\mathbf{s}_{\text{PD}})|\mathbf{s}_{\text{PD}}|^{\kappa}$ where $\kappa \rightarrow 0$ increases the compression rate and $\text{sgn}(\cdot)$ returns the sign of the argument.

\begin{figure*}
  \centering
  \includegraphics[width=7.0in]{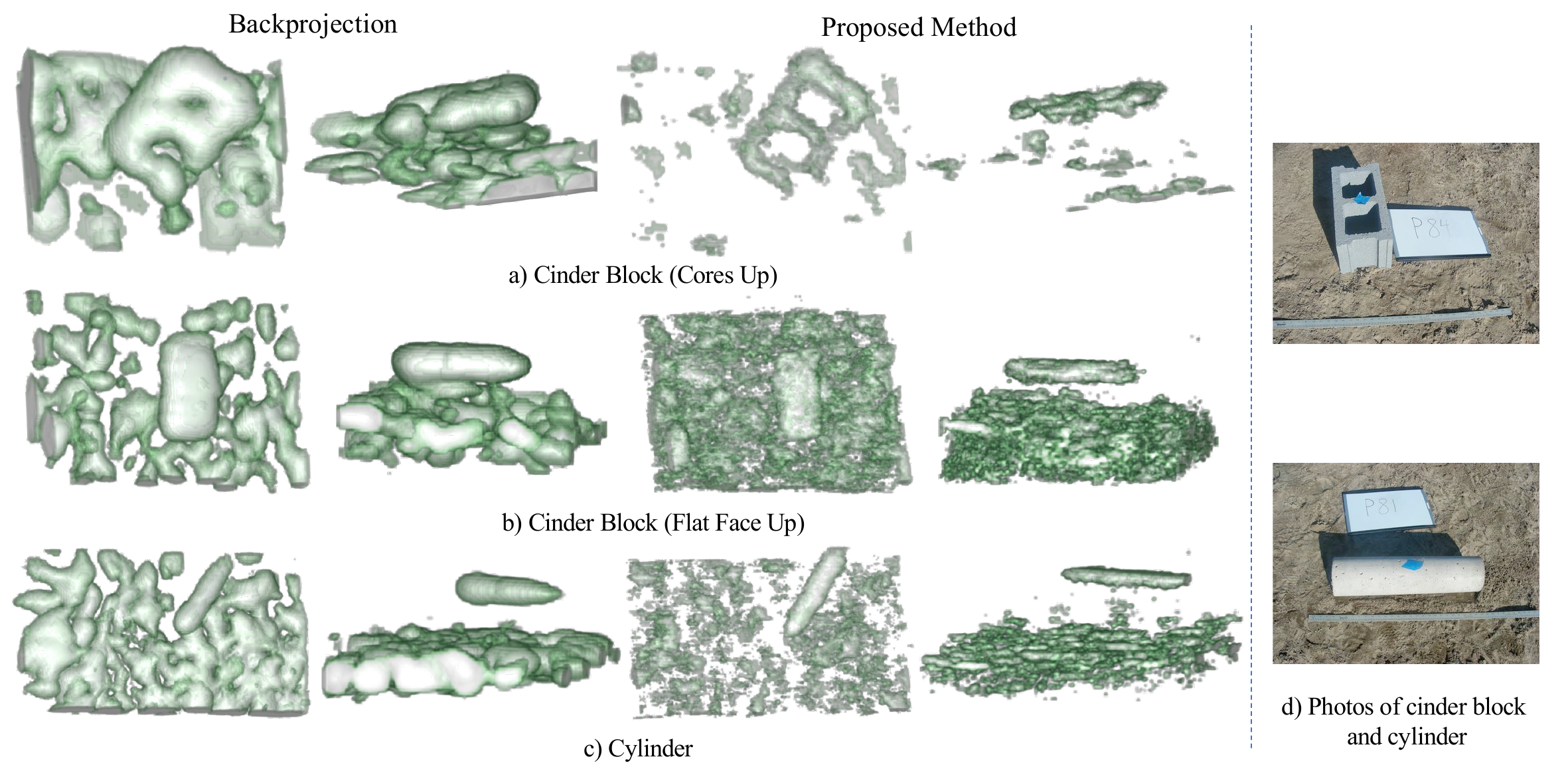}
  \caption{Two views of reconstructions with backprojection (left) and our proposed method (right) using in-water SAS data captured with SVSS. Our method recovers sharper object geometries, for example around the cinder block cores. Objects appear flatter in our reconstructions because of our pulse deconvolution step compressing measurements more than matched filtering. SVSS hardware photos courtesy of~\cite{svss-tech-report}.}
  \label{fig:svss-all-results}
\end{figure*}

In Fig.~\ref{fig:svss-all-results}, we show reconstructions of three targets of interest along an SVSS track. Note that we show a 2D maximum intensity projection (MIP) of an entire track in Fig.~\ref{fig:teaser}. Overall, our reconstructions appear sharper and with flatter side-profiles, because we better compress the waveforms during pulse deconvolution. In some cases, the object geometry appears more well-defined using our method, for example along the square cutouts in the cinder block. We also measure quantitative dimensions of the reconstructed cinder blocks (see supplemental material), and find that our method and backprojection are roughly equivalent in estimating metric lengths. 



\section{Discussion}
\label{sec:discussion}

We design an analysis-by-synthesis optimization for reconstructing SAS scenes, which yields several advantages to traditional approaches, including the ability to incorporate scene priors. We demonstrate the usefulness of this ability in Fig.~\ref{fig:undersampled-results}, where we achieve far better reconstructions in the undersampled regimes of helical and sparse-view collection geometries. Additionally, our choice of a differentiable forward model influences image formation in interesting ways. For example, our inclusion of the Lambertian scattering model drastically improves our reconstructions, as shown in Fig.~\ref{fig:ablation-normals}. These scattering constraints are easier to implement in our framework than with backprojection. We believe this work provides a promising avenue for future investigations into additional physics-based scattering models that enhance SAS reconstruction accuracy. 


While existing acoustic renderers for simulating SAS measurements exist, we highlight the challenges of designing an efficient differentiable forward model for SAS compatible with analysis-by-synthesis optimization. In particular, we discuss the burdensome sampling requirements for integrating spherically propagating acoustic sources. We addressed this challenge in two ways. First, we proposed a pulse deconvolution step that yields waveforms with energy distributed among sparse time bins. In Section~\ref{sec:ellipsoid} (and derived in the supplemental material), we show that synthesis of each of these time bins equates to integrating the scene points that lie on the surfaces of ellipsoids. Second, we implement importance sampling in range and direction, and compute return rays only at the expected depth to make our optimization tractable for our 3D reconstructions.



The performance of our method contrasts with backprojection's in meaningful ways. First, our method is less sensitive to transmit waveform bandwidth as shown in Fig.~\ref{fig:airsas-results} when comparing the $\Delta f = 20$ kHz and $\Delta f = 5$ kHz reconstructions. As bandwidth is expensive (i.e., expensive transducers), these results demonstrate the potential for high-quality SAS imaging with inexpensive hardware. Second, our method typically recovers more accurate 3D shape details than backprojection. This is due to our pulse deconvolution step compressing the waveforms for accurate localization, and properly modeling occlusion, surface normals, and Lambertian scattering in our forward model.  





\subsection{Limitations and Future Work}
There are disadvantages to analysis-by-synthesis frameworks applied to SAS reconstruction. First, our optimization is significantly slower than backprojection. Our reconstructions take up to 1-2 hours to complete, whereas backprojection can reconstruct scenes in minutes. Practically this implies a trade-off for the choice of reconstruction: backprojection may be most useful when surveying large underwater regions, whereas our method can be applied to enhance visual details to regions of interest identified in the backprojected imagery. 

The second disadvantage of analysis-by-synthesis is that our reconstruction quality is limited by the accuracy of our forward model. Using a neural network in our pipeline can help compensate for forward model inaccuracies~\cite{xie2022neural, lucas2018using}, but this is true only to an extent. We observe that our model falls short of reconstructing effects ignored by our forward model, such as elastic scattering. Elastic scattering describes the phenomena where an insonified target stores then radiates acoustic energy~\cite{al2021acoustic}. Elastic scattered energy can be seen as non-zero energy that appears to radiate downward from the cylinder approximately centered on the backprojected 3D reconstruction slice in Fig.~\ref{fig:teaser}. The presence of this energy is useful for target classification and detection~\cite{brown2019simulation}. While our proposed method predicts sharper object boundaries, this energy is notably absent in our reconstruction. This makes sense since our forward model has no mechanism to handle non-linear acoustics. An interesting future direction is to incorporate physics-based models that handle non-linear acoustics in our formulation. 

There are interesting directions for future work related to accounting for uncertainty in our forward model. First, future work may find our method useful for solving joint unknown problems. For example, there is typically uncertainty in the SAS platform's position with respect to the scene~\cite{hayes2009synthetic}. Future work may investigate using our analysis-by-synthesis framework to jointly solve for the platform position and the scene. Second, surrogate modeling may be useful for approximating inefficient and non-differentiable forward models with neural networks~\cite{sun2019review}. 

We now address some limitations with our pulse deconvolution step. In Fig.~\ref{fig:quad-ablation}, we demonstrate the importance of this step, as using the matched filtered waveforms as input to neural backprojection drastically underperforms using the deconvolved pulse waveforms. However, deconvolution is an ill-posed inverse problem that is sensitive to noise, making the performance of this step depend on sparsity and smoothness hyperparameters. Thus, pulse deconvolution requires user input to select hyperparameters that maximize deconvolution performance. 
Future work may seek ways to robustly deconvolve the waveforms that minimize the need for hyperparameter tuning. 


Finally, we mention that our deconvolution method and subsequent neural backprojection are done at the carrier frequency $f_c$, rather than at the baseband spectrum. Since we are operating at a relatively low carrier frequency ($f_c = 20$ and $f_c = 27.5$ kHz for AirSAS and SVSS, respectively), relative to our sampling rate $F_s = 100$ kHZ, this is a non-issue. However, adapting this method to radar or even higher frequency SAS may require adapting the method to baseband signals. Empirically, we observed that trivially adapting our method to basebanded measurements results in worse reconstructions and therefore, it requires further investigation.

\subsection{Conclusion}
This work presents a reconstruction technique for SAS that we show outperforms traditional image formation in a number of settings. Importantly, we demonstrate that the method scales to in-water SAS data captured from field surveys. We believe this work to be an important step in advancing SAS imaging since it provides a framework for incorporating physics-based knowledge and custom priors into SAS image formation. More broadly, as this work demonstrates an impactful application of neural rendering for SAS, we believe it opens new possibilities for other synthetic aperture and coherent imaging fields like radar and ultrasound. 


\begin{acks}
This material was supported by ONR grant N00014-23-1-2406 as well as SERDP Contract No. W912HQ21P0055: Project MR21-1334. A. Reed was supported by a DoD NDSEG Fellowship. T. Blanford was supported by ONR grant N00014-22-1-2607. The authors acknowledge Research Computing at Arizona State University for providing GPU resources. We thank Christopher Eadie for 3D printing the AirSAS objects.
\end{acks}

\bibliographystyle{ACM-Reference-Format}
\bibliography{sample-base}
\end{document}


\title{Supplemental Material: Neural Volumetric Reconstruction for Coherent Synthetic Aperture Sonar}

\maketitle

In this supplemental material, we present a derivation of our analytic forward model, additional results, and ablation studies to justify our method presented in the main paper. We also encourage the reader to view the supplemental videos of revolving 3D reconstructions as well as the code and data\footnote{https://awreed.github.io/Neural-Volumetric-Reconstruction-for-Coherent-SAS/}.

\section{Derivation of Analytic Forward Model with Pulse Deconvolved Measurements}
We first provide the full derivation of our analytic forward model to synthesize pulse deconvolved waveforms. Our starting point is the forward model given by point-based sonar scattering~\cite{brown2017modeling,brown2017point}:

\begin{equation}
\label{eq:A1}
    \mathbf{s}(t) = \int\displaylimits_{\mathcal{X}} \frac{b_T(\mathbf{x})b_R(\mathbf{x})T(\mathbf{o}_T, \mathbf{x})T(\mathbf{o}_R, \mathbf{x})}{2\pi R_T R_R}\sigma(\mathbf{x})p\left(t - \frac{R_t + R_R}{c}\right)\mathrm{d}\mathbf{x}.
\end{equation}

We redefine the terms in Eq.~\ref{eq:A1} for completeness. The real-valued sonar measurements in time are denoted by $\mathbf{s}(t)$. The set of all scene points is given by $\mathcal{X}$, and an individual point is given by $\mathbf{x}$. $\mathbf{o}_T$ and $\mathbf{o}_R$ denote the transmitter and receiver positions, respectively. The transmitter and receiver directivity functions are given by $b_T(\mathbf{x})$ and $b_R(\mathbf{x})$. The transmission probability from the transmitter to a scene point is given by $T(\mathbf{o}_T, \mathbf{x})$; its probability of return back to the receiver is given by $T(\mathbf{o}_R, \mathbf{x})$. The distance from the transmitter to a scene point is given by $R_T$; $R_R$ denotes the distance from the point back to the receiver. The scene scatterer amplitudes are given by the function $\sigma(\cdot)$. The transmit pulse in time is defined by $p(t)$ and is by the time-of-flight to a scene point using the sound speed, $c$.

For the sake of readability, we define the term $$K(\mathbf{x},\mathbf{o}_R,\mathbf{o}_T) = \frac{b_T(\mathbf{x})b_R(\mathbf{x})T(\mathbf{o}_T, \mathbf{x})T(\mathbf{o}_R, \mathbf{x})}{2\pi R_T R_R}$$ for the integral in this equation and subsequent derivations. We also note that $\sigma$ can be any scattering model including the Lambertian model $L(\sigma)$ used in the main paper. 

If we perform cross-correlation on the equation above, we would compute \begin{equation}
\label{eq:A2}
    \mathbf{s}(t) = \mathbf{s}(t) \ast_t p^{*}(-t).
\end{equation}

Combining Eq.~\eqref{eq:A1} and~\eqref{eq:A2} and using the linearity properties of the convolution operator, we get the following equation:
\begin{equation}
\label{eq:A3}
    \mathbf{s}(t) = \int\displaylimits_{\mathcal{X}} K(\mathbf{x},\mathbf{o}_R,\mathbf{o}_T)\cdot \sigma(\mathbf{x})\cdot p\left(t - \frac{R_t + R_R}{c}\right) \ast_t p^{*}(-t) \mathrm{d}\mathbf{x}.
\end{equation}

Constructing the analytic signal for the measurement given in Eq.~\eqref{eq:A3}, we get that:

\begin{equation}
\label{eq:A4}
    \mathbf{\widehat{s}} = \mathbf{s} + j\mathcal{H}(\mathbf{s}),
\end{equation}

\begin{multline}
\label{eq:A5}
    \mathbf{\widehat{s}} = \int\displaylimits_{\mathcal{X}} K(\mathbf{x},\mathbf{o}_R,\mathbf{o}_T)\cdot \sigma(\mathbf{x})\cdot p\left(t - \frac{R_t + R_R}{c}\right) \ast_t p^{*}(-t) \mathrm{d}\mathbf{x} 
    \\ + j\mathcal{H}\left( \int\displaylimits_{\mathcal{X}} K(\mathbf{x},\mathbf{o}_R,\mathbf{o}_T)\cdot \sigma(\mathbf{x})\cdot p\left(t - \frac{R_t + R_R}{c}\right) \ast_t p^{*}(-t) \mathrm{d}\mathbf{x} \right).
    \end{multline}

Using the linearity of the Hilbert transform and regrouping terms, 

\begin{equation}
\label{eq:A6}
\mathbf{\widehat{s}} = \int\displaylimits_{\mathcal{X}} K(\mathbf{x},\mathbf{o}_R,\mathbf{o}_T)\cdot \sigma(\mathbf{x})\cdot \widehat{P}(t) \mathrm{d}\mathbf{x},
\end{equation}

where $$ \widehat{P}(t) = \left(p\left(t - \frac{R_t + R_R}{c}\right) \ast_t p^{*}(-t)\right) + j \mathcal{H}\left(p\left(t - \frac{R_t + R_R}{c}\right) \ast_t p^{*}(-t)\right).$$ Here, the term $\widehat{P}$ represents the analytic signal of the cross-correlated pulse. 

We now perform a common modeling trick of assuming the point scattering field is complex, i.e. $\netout = \sigma \cdot \widehat{P}$, and have our method estimate these values instead. Note that a similar trick was performed by Reed et al.~\cite{SINR} for 2D SAS image deconvolution.

\begin{align}
   \widehat{\mathbf{s'}_{\text{PD}}}\left(t=\frac{R_T+R_R}{c}\right) =  \int_{\mathcal{X}} \frac{b_T(\mathbf{x})T(\mathbf{o}_T, \mathbf{x})T(\mathbf{x}, \mathbf{o}_R)}{2\pi R_T R_R}\netout(\mathbf{x})\mathrm{d}\mathbf{x}, 
   \label{eq:xfinal}
\end{align}

\paragraph{Ideal Pulse Deconvolution.} If we assume that our pulse deconvolution described in Section 5 is ideal, then $p\left(t - \frac{R_t + R_R}{c}\right) \ast_t p^{*}(-t) = \delta\left( t - \frac{R_t + R_R}{c}\right)$ is an ideal delta function. Then we can simplify Eq.~\eqref{eq:A6} as follows:

\begin{equation}
\mathbf{\widehat{s}}(t) = \int\displaylimits_{\mathcal{X}} K(\mathbf{x},\mathbf{o}_R,\mathbf{o}_T)\cdot \sigma(\mathbf{x})\cdot \widehat{\delta}\left( t - \frac{R_t + R_R}{c}\right) \mathrm{d}\mathbf{x}.
\end{equation}

Computing the analytic signal of the delta function using the fact that $\mathcal{H}(\delta(t)) = \frac{1}{\pi t}$, we get:
\begin{equation}
\widehat{\delta}\left( t - \frac{R_t + R_R}{c}\right) = \delta\left( t - \frac{R_t + R_R}{c}\right) + \frac{j}{\pi \left(t-\frac{R_t + R_R}{c}\right)}
\label{eq:deltahat}
\end{equation}

Now we can use the fact that $\widehat{P} = \widehat{\delta}\left(t-\frac{R_t + R_R}{c}\right)$ has most of its energy at the time-of-flight $t = \frac{R_t + R_R}{c}$. 
As shown in Fig. 5 in the main paper, if we assume a one-bounce reflection model, then the set of points with a constant time-of-flight $t=\frac{R_t + R_R}{c}$ describe an ellipsoid with a semi-major axis of length $r = c\cdot t/2$, where $c$ is the sound speed. Specifically, these points define the ellipsoid:
\begin{equation}
    \frac{x^2}{a(r)^2} + \frac{y^2}{b(r)^2} + \frac{z^2}{c(r)^2} - 1 = 0, 
    \label{eq:ellipse-equation}
\end{equation}
where transmit $\mathbf{o}_T$ and receive $\mathbf{o}_R$ elements are separated by distance $d$ and the ellipsoid axes are, 
\begin{equation}
    a(r) = r, b(r)=\sqrt{(r)^2 - (d/2)^2}, c(r) = b(r).
\end{equation}
We let $\mathbf{E}_{r}$ be the set of $\mathbf{x}$ points on the surface of the ellipsoid defined by range $r$. Thus, we can approximate Eq.~\eqref{eq:xfinal} since Eq.~\eqref{eq:deltahat} has most of its energy near $t=\frac{R_t + R_R}{c}$, and thus restrict the domain of integration to $\mathbf{E}_r$:
\begin{equation}
\label{eq:Adelta}
\mathbf{\widehat{s}}\left(t=\frac{R_t + R_R}{c}\right) \approx \int\displaylimits_{\mathbf{E}_{r}} K(\mathbf{x},\mathbf{o}_R,\mathbf{o}_T)\cdot \netout(\mathbf{x})  \mathrm{d}\mathbf{x}.
\end{equation}

Equation ~\eqref{eq:Adelta} is the final forward model that we utilize in our main method to synthesize $\mathbf{\widehat{s}(t)}$ that is optimized against the real measurements from the sonar. In implementation, we omit the spherical spreading term $\frac{1}{2\pi R_T R_R}$ in the equation because its effects are negligeable for our relatively small scene sizes. We note that this term is commonly ommitted in time-domain beamformer implementations~\cite{hayes1992broad}. Limitations of the forward model include not modeling diffraction, a single bounce assumption, and difficulty recovering elastic scattering effects. We discuss these limitations in more detail in the main paper.






\section{Time of Flight Renderer Details}

In this section, we describe the details of the renderer implementation used to simulated SAS measurements. We use a transient renderer that is bootstrapped to the optical renderer~\cite{kim2021fast}, which uses the GPU ray tracing library, OptiX. The transient renderer exploits a CUDA atomic operation on top of the optical renderer to render transient signals.
We render each object at 54000 different camera poses, each pose corresponding to a measurement position from AirSAS, and the transient signal sampled at the AirSAS sampling frequency of $100$~kHz.

\begin{figure}
  \centering
  \includegraphics[width=3.35in]{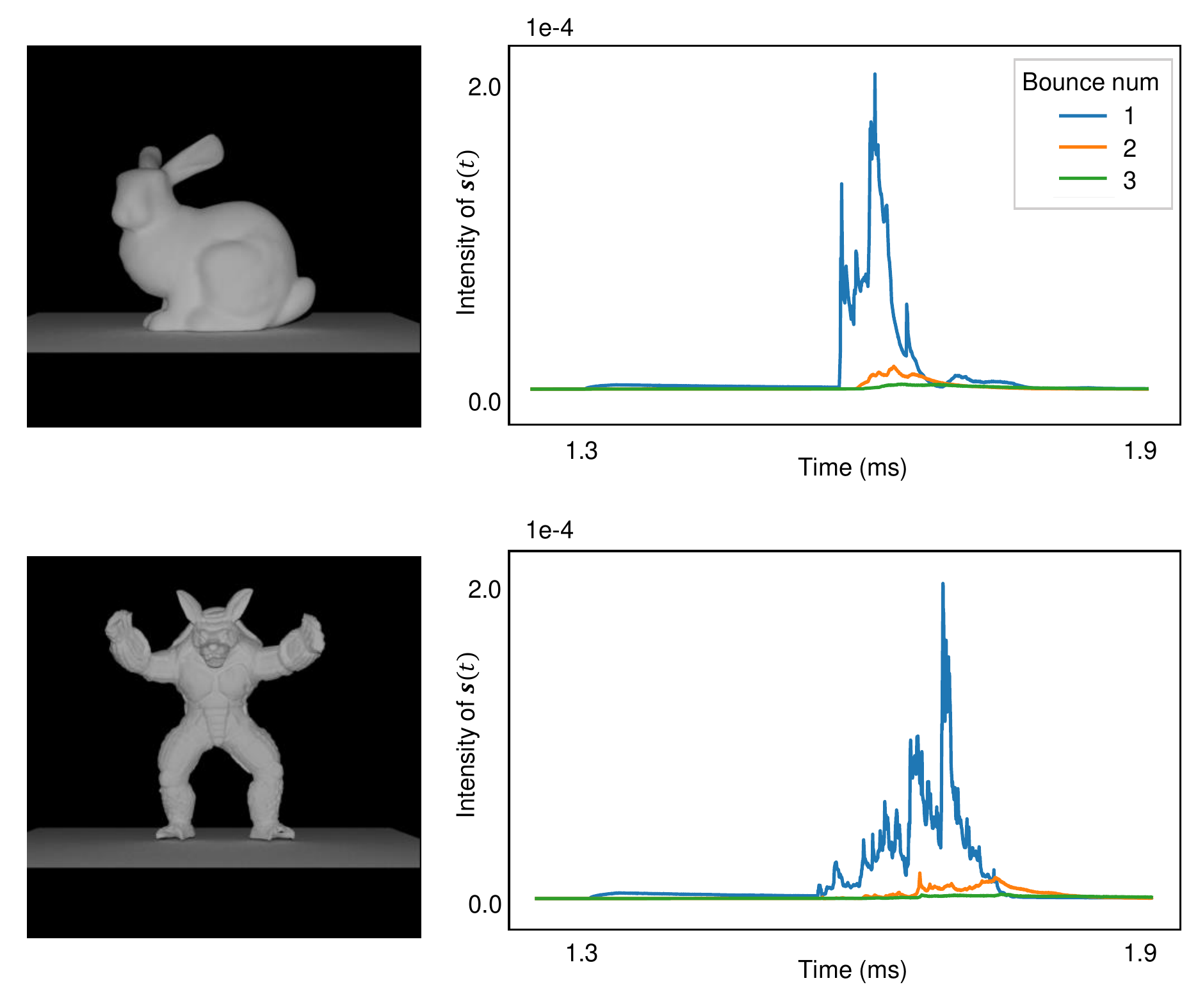}
  \caption{Rendered image and transient signal from one sensor location for the bunny and armadillo objects.}
\label{fig:renderer_image_and_signal}
\end{figure}

Examples of rendered AirSAS scenes and the corresponding transient signals from one sensor location can be found in Fig.~\ref{fig:renderer_image_and_signal}. We add a plane below the objects to emulate the physical AirSAS setup, which is confined with an anechoic chamber that is padded everywhere except the ground. Accumulating rendered rays by their bounce number shows that the intensity of the single-bounced signal is dominant.


\section{Additional Simulated Results}



\begin{figure*}
  \centering
  \includegraphics[width=5.5in]{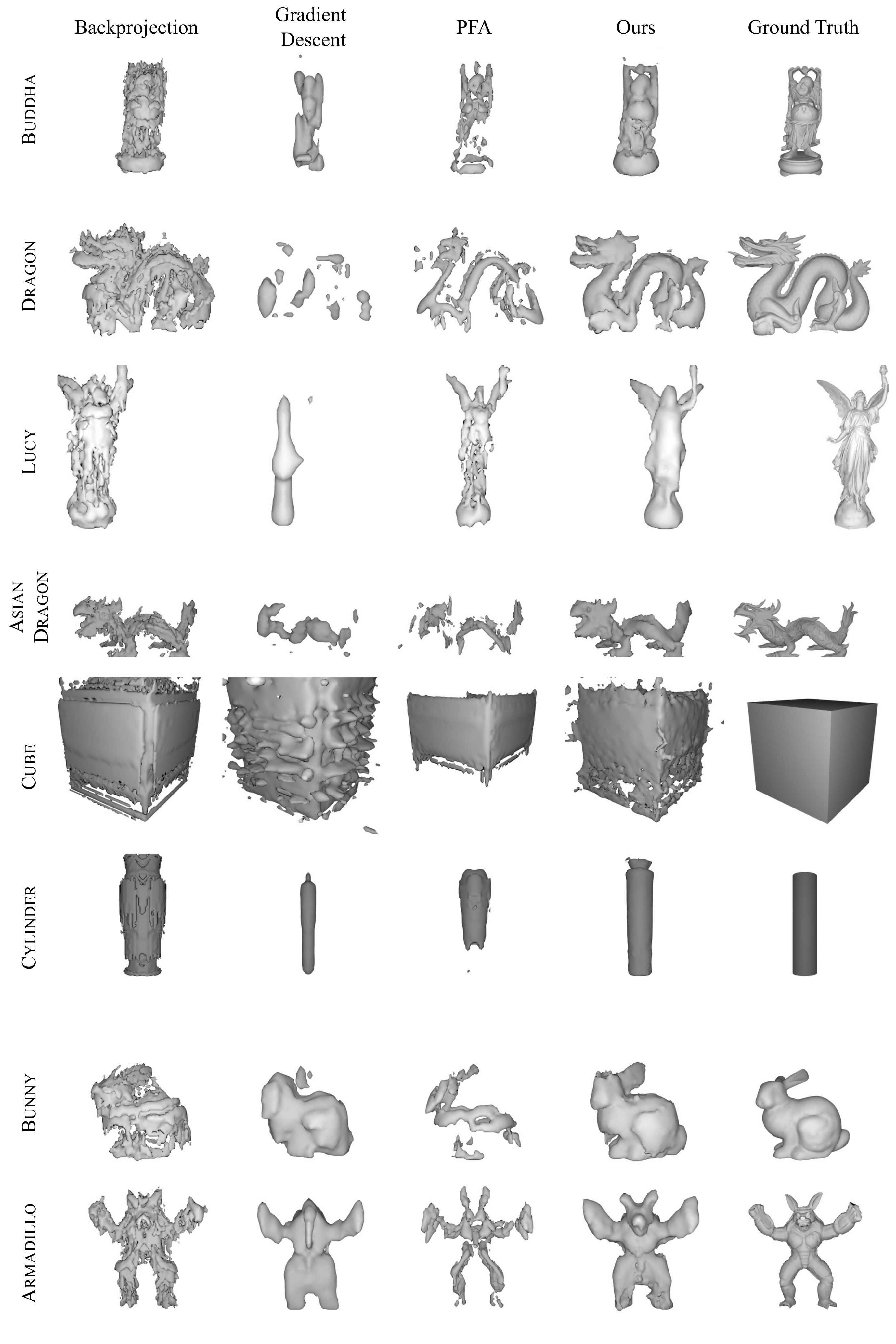}
  \caption{Rendered meshes using different methods. The simulation was performed under bandwidth 20k with a noise level of 20db. We observe that our technique consistently performs better than traditional techniques. }
  \label{fig:more_meshes}
\end{figure*}

\begin{figure*}
  \centering
  \includegraphics[width=5.5in]{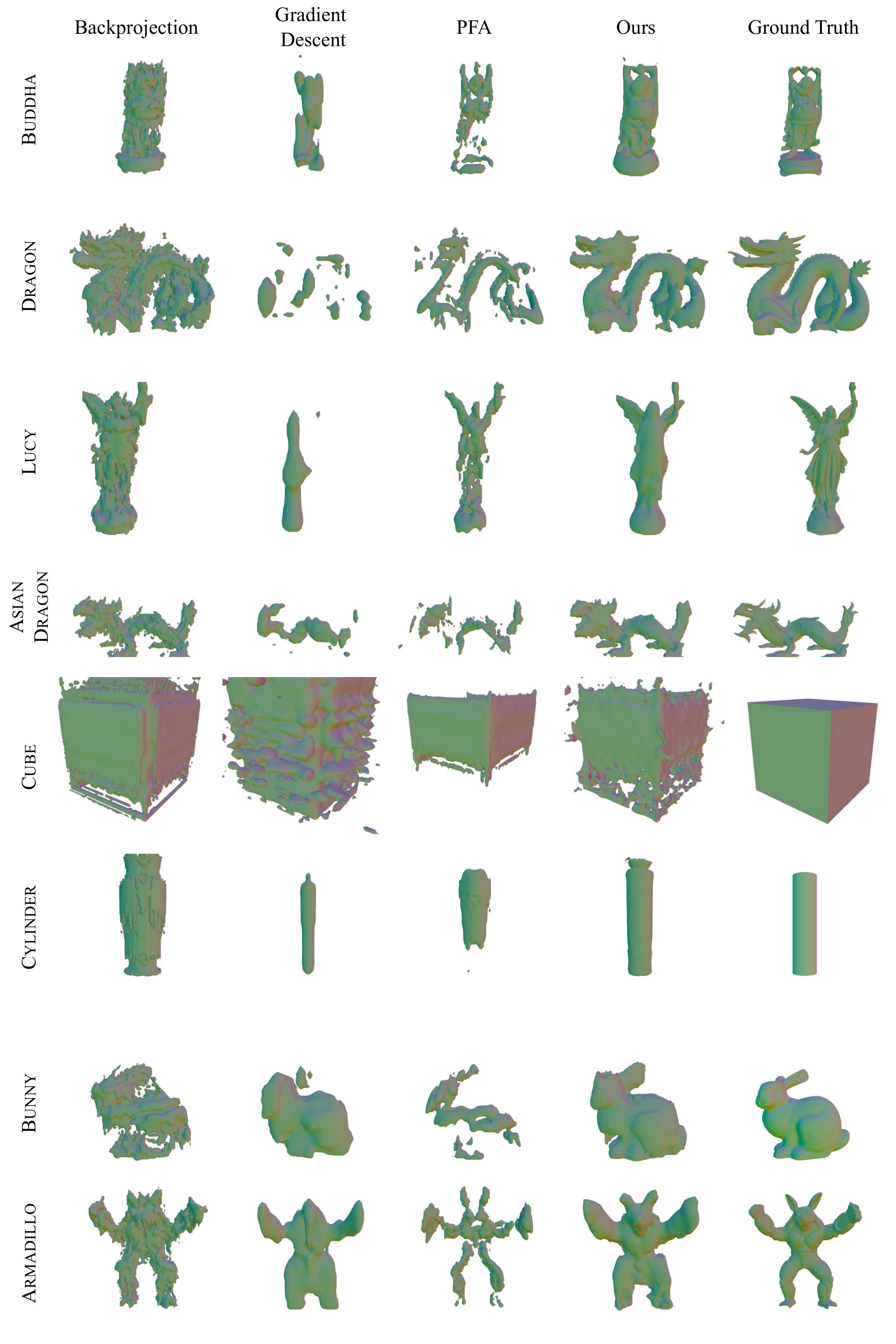}
  \caption{Normal of rendered meshes using different methods. The simulation was performed under bandwidth 20k with the noise level of 20db. Our technique reconstructs normal maps that are more similar to the ground truth than previous techniques. }
  \label{fig:more_meshes_normal}
\end{figure*}

We show the results for simulated data for both shape reconstruction (Fig.~\ref{fig:more_meshes}) and surface normal reconstruction (Fig.~\ref{fig:more_meshes_normal}). All simulations assume $\Delta f = 20 kHz$ and a signal-to-noise ratio of 20dB. The quantitative metrics from these eight meshes are summarized in Table 2 in the main paper. Our method can consistently achieve high-quality reconstructions more faithful to the ground truth meshes in both shape and surface normal distribution.

\section{Ablation Studies}
\begin{figure}
  \centering
  \includegraphics[width=3.35in]{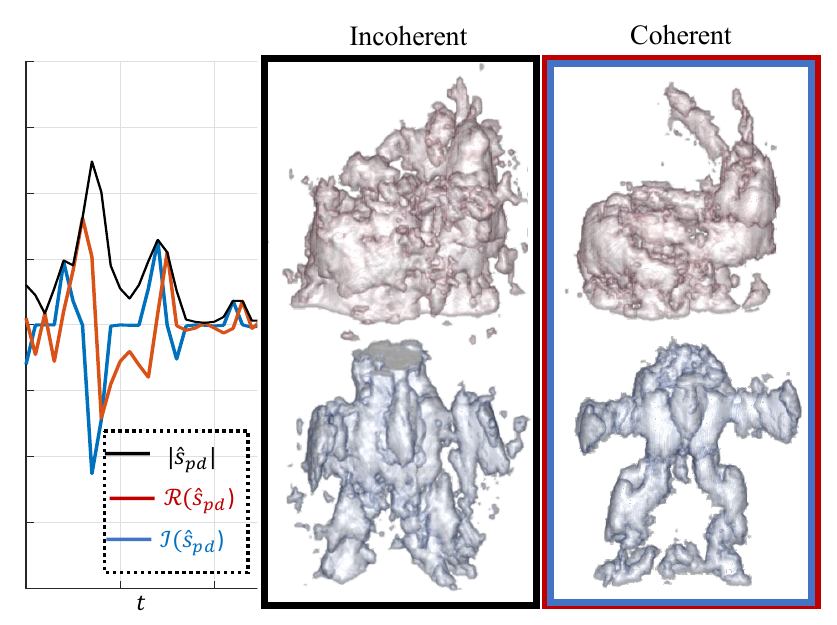}
  \caption{Reconstructions from our method using incoherent and coherent processing of $\Delta f = 20$ kHz AirSAS bunny and armadillo. Incoherent processing integrates the envelope (black line) of measurements; coherent processing integrates the complex analytic signal (red and blue lines) of measurements. Clearly, coherent processing is necessary for maximizing performance.}
  \label{fig:incoherent-vs-coherent-v2}
\end{figure}

\paragraph{Coherent versus incoherent reconstruction:} In the main paper, we describe how we compute our loss between the complex-valued analytic measurements and network synthesized measurements. In particular, our network outputs a complex-valued scene $\netout$ to synthesize these complex measurements. In contrast, some sonar reconstructions (e.g. forward-look) leverage an incoherent reconstruction where only the magnitude of the signal is used~\cite{kim2005mosaicing, kim2007enhanced}. To compare coherent versus incoherent reconstruction, we configured our network to output a real-valued scene to synthesize the magnitude of measurements directly for incoherent reconstruction, which we show in Fig.~\ref{fig:incoherent-vs-coherent-v2}. The left column of the figure shows the complex analytic signal (red and blue curves) used in the coherent reconstruction and their magnitude (black curve) which is used for the incoherent reconstruction. The performance disparity aligns with SAS processing fundamentals that measurements should be coherently processed to enhance resolution. 

\begin{figure}
  \centering
  \includegraphics[width=2.75in]{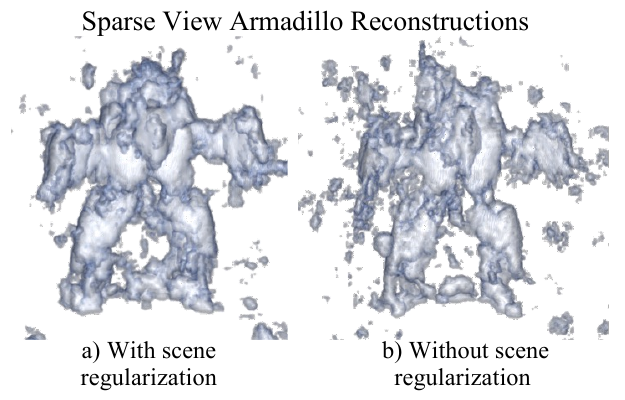}
  \caption{Effect of Prior: (Ablations using sparse view AirSAS measurements
of the 20 kHz armadillo). We show a sparse sampled armadillo reconstruction with
and without scene priors. Scene priors are especially helpful in the sparse
and helical sampling cases to attenuate undersampling artifacts.}
  \label{fig:priors}
\end{figure}

\paragraph{Effect of regularization terms:} Finally, we studied the impact of our scene regularizations in Fig.~\ref{fig:priors}. The neural backprojection smoothness and sparsity priors are sometimes useful in certain conditions. We find that they are especially helpful for sparse view reconstructions, as they attenuate streaking artifacts that occur due to having limited angular measurements. In the figure, we show that using the smoothness and sparsity priors attenuates noise in the scene and enables a more accurate reconstruction of the armadillo object.

\begin{figure}
  \centering
  \includegraphics[width=3.4in]{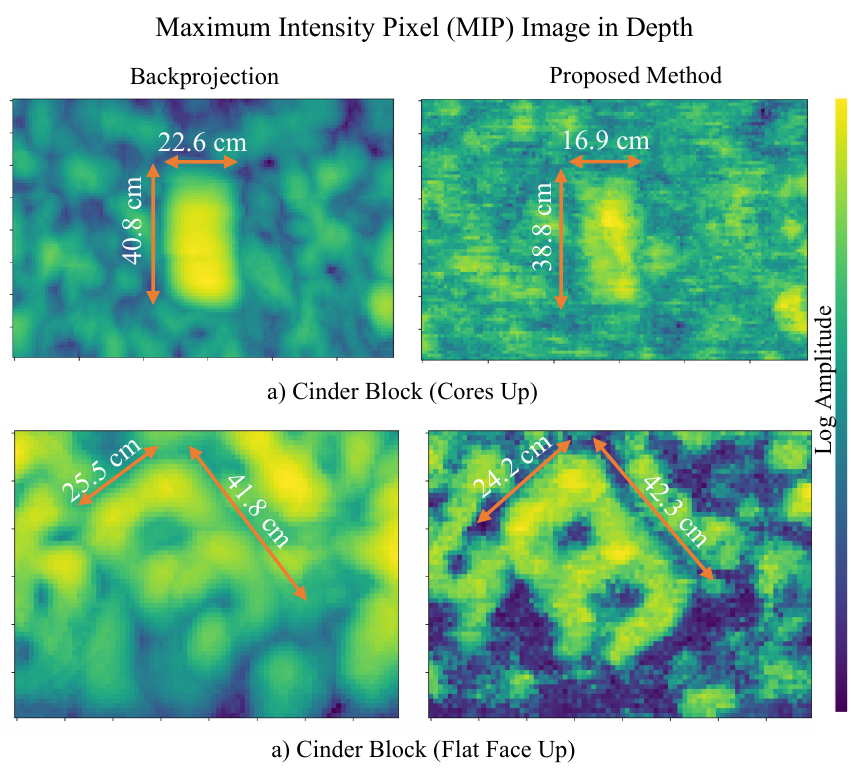}
  \caption{MIP images of the (a) cinder block with cores up and (b) cinder block face up. The ground truth dimensions for the cinder block are $20 \times 40$ centimeters (cm). Our proposed method performs comparably to backprojection in the accuracy of the reconstructed dimensions. Note that each image pixel is $1 \times 1 \text{cm}$.}
  \label{fig:svss-quant}
\end{figure}

\section{Additional SVSS Results}

\paragraph{SVSS Quantitative:} In Figure~\ref{fig:svss-quant}, we show the maximum intensity projection (MIP) in the depth dimension for the same cinder block targets shown in the main text. While measuring the dimensions of the targets from the MIP is an approximation since we are actually measuring the projection of pixels onto a plane, we show that our method retains most of the accuracy of backprojection, typically reconstructing the target dimensions typically within a few centimeters of ground truth. Our method tends to estimate a size slightly smaller than ground truth, whereas backprojection estimates slightly larger. This is perhaps because of our pulse deconvolution step compressing measurements in range more than matched filtering. 

\begin{figure*}[h!]
  \centering
  \includegraphics[width=7.0in]{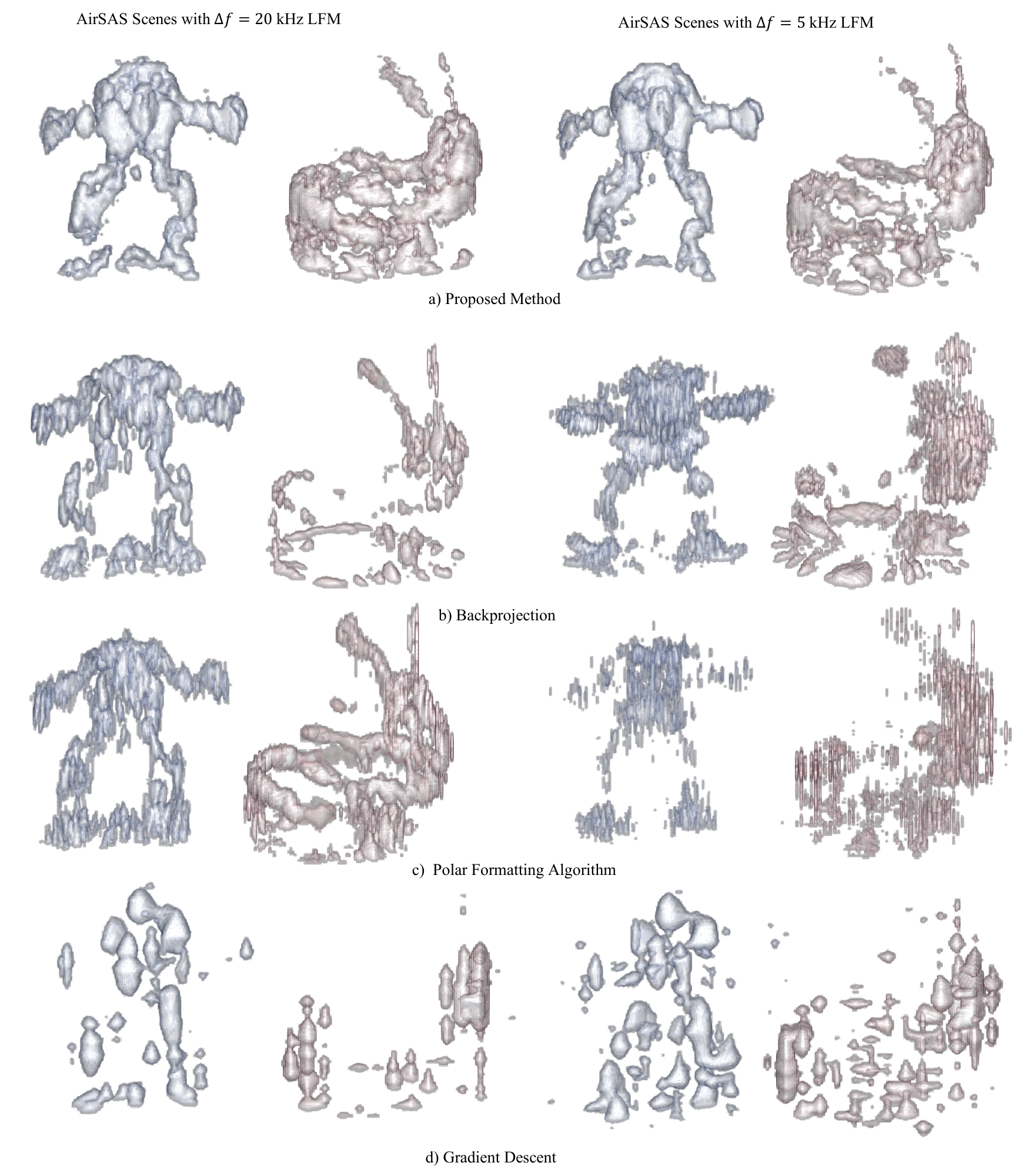}
  \caption{Higher threshold used to visualize reconstructions of AirSAS data captured with relatively high ($\Delta f = 20 $ kHz) and low ($\Delta f = 5 $ kHz) bandwidth LFMs. Our method demonstrates more consistent performance across waveform bandwidth compared to backprojection, the polar formatting algorithm, and gradient descent.}
  \label{fig:airsas-results-high-thresh}
\end{figure*}

\begin{figure*}[h!]
  \centering
  \includegraphics[width=7.0in]{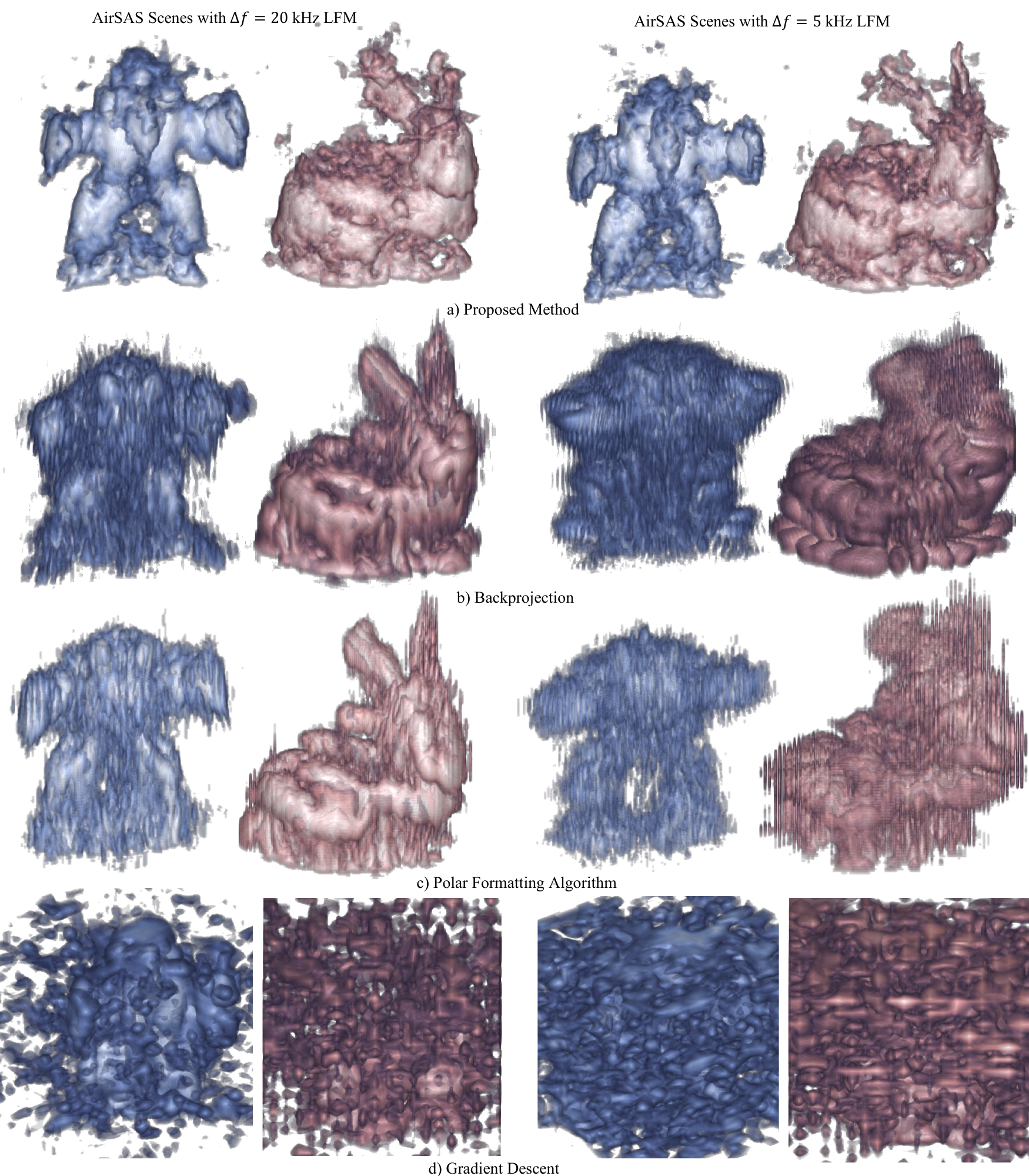}
  \caption{Lower threshold used to visualize reconstructions of AirSAS data captured with relatively high ($\Delta f = 20 $ kHz) and low ($\Delta f = 5 $ kHz) bandwidth LFMs. Our method demonstrates more consistent performance across waveform bandwidth compared to backprojection, the polar formatting algorithm, and gradient descent.}
  \label{fig:airsas-results-low-thresh}
\end{figure*}

\section{Additional AirSAS Results}

\paragraph{Effects of rendering threshold parameter:}
In the main text, we normalize the magnitude of AirSAS reconstructions between $[0-1]$ and use a threshold of $0.2$ for visualization (i.e. set all magnitudes < 0.2 = 0). In Fig.~\ref{fig:airsas-results-high-thresh} and Fig.~\ref{fig:airsas-results-low-thresh}, we render the AirSAS results from the main text at higher (0.3) and lower (0.05) thresholds, respectively. While the threshold affects the qualitative appearance of all methods, we observe that our method retains better retains object geometry across all thresholds when compared with backprojection and gradient descent. 

\clearpage

\bibliographystyle{ACM-Reference-Format}
\bibliography{sample-base}